\documentclass[]{interact}
\usepackage{amssymb, amsmath, amsthm, mathrsfs}
\usepackage{mathtools}
\usepackage{bm}
\graphicspath{{Pictures/}}
\usepackage{hyperref}
\usepackage{xcolor}

\usepackage{epstopdf}
\usepackage[caption=false]{subfig}

\usepackage[numbers,sort&compress]{natbib}
\bibpunct[, ]{[}{]}{,}{n}{,}{,}
\renewcommand\bibfont{\fontsize{10}{12}\selectfont}
\makeatletter
\def\NAT@def@citea{\def\@citea{\NAT@separator}}
\makeatother

\theoremstyle{plain}

\theoremstyle{definition}

\theoremstyle{remark}

\newcommand*\Laplace{\mathop{}\!\mathbin\bigtriangleup}
\newcommand{\Qvec}{\mathbf{Q}}

\newcommand{\nvec}{\mathbf{n}}
\newcommand{\I}{\mathbf{I}}

\usepackage{cleveref}

\begin{document}

\articletype{ARTICLE TEMPLATE}

\title{Stochastic effects on solution landscapes for nematic liquid crystals}

\author{
\name{J.~L. Dalby\textsuperscript{a}\thanks{CONTACT J.~L. Dalby. Email: james.dalby@strath.ac.uk}, A. Majumdar\textsuperscript{a} \thanks{apala.majumdar@strath.ac.uk}, Y. Wu\textsuperscript{a}, A.~K. Dond\textsuperscript{b}} 
\affil{\textsuperscript{a}Department of Mathematics and Statistics, University of Strathclyde, Glasgow, UK; \textsuperscript{b} School of Mathematics, IISER Thiruvananthapuram, Vithura, India} 
}

\maketitle

\begin{abstract}
 We study the effects of additive and multiplicative noise on the solution landscape of nematic liquid crystals confined to a square domain within the Landau-de Gennes framework, as well as the impact of additive noise on the symmetric radial hedgehog solution for nematic droplets. The introduction of random noise can be used to capture material uncertainties and imperfections, which are always present in physical systems. We implement random noise in our framework by introducing a $Q$-Wiener stochastic process to the governing differential equations. On the square, the solution landscape for the deterministic problem is well understood, enabling us to compare and contrast the deterministic predictions and the stochastic predictions, while we demonstrate that the symmetry of the radial hedgehog solution can be violated by noise. 
 This approach of introducing noise to deterministic equations can be used to test the robustness and validity of predictions from deterministic liquid crystal models, which essentially capture idealised situations. 
\end{abstract}


\section{Introduction}
Liquid crystals are classical examples of partially ordered materials that have a degree of positional and/or orientational order \cite{deGennes}. The simplest, and most common liquid crystal phase, is the nematic phase, for which the constituent rod-like molecules typically exhibit a degree of long-range orientational ordering i.e. the nematic phase has distinguished material directions referred to as \emph{directors}. From a modelling perspective, liquid crystals are studied almost exclusively via deterministic models, such as the Oseen-Frank, Ericksen-Leslie and Landau-de Gennes (LdG) continuum theories, as well as various molecular models. However, their stochastic counterpart i.e., introducing random noise to the governing differential equations, has received less attention despite the fact they have the potential to account for material imperfections and uncertainties that will almost always be present in physical systems. 

We work with the LdG theory, wherein the state of nematic ordering is captured by the LdG $\mathbf{Q}$-tensor, which is a symmetric, traceless $3\times 3$ matrix \cite{deGennes}. The physically observable configurations are modelled by local or global minimisers of an appropriately defined LdG free energy, which is typically a nonlinear and non-convex functional of the LdG $\mathbf{Q}$-tensor and its derivatives. Mathematically, the energy minimisers are classical solutions of the associated system of Euler-Lagrange equations - a system of five nonlinear elliptic partial differential equations subject to appropriate boundary conditions. We study the effects of random noise on the solutions of the LdG Euler-Lagrange equations for two well known problems; (i) nematic liquid crystals confined to two-dimensional square domains \cite{canevari_majumdar_spicer2017,2D_landscape} and (ii) a spherical droplet of nematic liquid crystal \cite{gartland_hedgehog}. 
The noise appears in the form of one additional term in the differential equations, modelled by a $Q$-Wiener stochastic process \cite{into_book}. This term models random perturbations 
which can capture material imperfections and/or uncertainties in the material properties and experimental set-up. For example, the random noise could capture the effects of material inhomogeneities, thermal fluctuations, manufacturing imperfections and/or material defects. 
On square domains, we consider additive and multiplicative noise, that is noise which is independent of the unknown we are solving for and noise which does depend on the unknown we are solving for \cite{into_book}, while for the droplet, we consider additive noise only. In both cases, we observe that random noise has a pronounced effect on solution profiles for small domains, and has a symmetry-breaking effect on highly symmetric solutions. 


To the best of our knowledge, the introduction of stochastic terms into a LdG model has not been considered in detail, in the literature. 
In \cite{stochastic_nematic}, the authors look at a modified Ericksen-Leslie model with multiplicative noise. They prove various rigorous results including the existence of a weak solution, pathwise uniqueness of the solution in 2D settings and a maximum principle. In \cite{stochastic_nematic_LDP}, a large deviation principle is developed for the same modified Ericksen-Leslie model with multiplicative noise, which may be useful for the study of switching processes between distinct equilibria in liquid crystal systems. 
However, these papers do not focus on numerical experiments or specific model examples as we do, for which one can compare the deterministic predictions with their stochastic counterparts. 
Numerical experiments are always vitally important as they give us detailed structural information about observable equilibria, including their multiplicity, singular sets and numerical experiments can explore parameter regimes inaccessible to rigorous asymptotic analysis. 

With this in mind, we perform a numerical exploration of the stochastic solution landscape for nematic liquid crystals in the LdG framework, for two model problems which have been studied extensively in the deterministic case. On square domains with tangent boundary conditions, it is well-known that the Well Order Reconstruction Solution (WORS) with two diagonal defect lines is globally stable in certain LdG frameworks for small square domains, whereas large square domains are multistable in the sense that they support two stable diagonal and four stable distinct rotated solutions \cite{canevari_majumdar_spicer2017, wang_canevari_majumdar_2019}. The tangent boundary conditions require the nematic director (modelled by the eigenvector of the LdG $\mathbf{Q}$-tensor with the largest positive eigenvalue) to be tangent to the square edges, creating a mismatch or defects at the square vertices. The WORS has a perfectly symmetric profile, such that the square diagonals partition the square domain into four quadrants and the nematic director is approximately constant in each quadrant. The director is not defined along the two square diagonals, and hence, the two square diagonals are interpreted as defects of the WORS solution. The diagonal and rotated solutions can be distinguished by the director profiles near the square vertices, which are defects induced by the tangent boundary conditions, and as the name suggests, the nematic director is approximately aligned with one of the square diagonals for the diagonal solutions. With random noise, we compute the probabilities of observing different solutions and relate stability to the probability of observation, in a non-rigorous manner. The introduction of random noise immediately kills the symmetry of the WORS and typically shrinks the domain of stability of WORS-like solutions. The impact of random noise on the diagonal and rotated solutions is less pronounced, and as such, we deduce that random noise has a more pronounced effect on the solution profiles for small domains or for solutions which have small domains of stability.  The second example concerns spherical droplets filled with nematic liquid crystals, with strong radial anchoring or homeotropic boundary conditions. The radial hedgehog solution has a perfectly radial nematic director with a single isolated point defect at the droplet centre, and the radial hedgehog solution is globally stable for sufficiently small droplets or for relatively high temperatures \cite{majumdar_2012_radialhedgehog}. For larger droplets, the radial hedgehog solution loses stability with respect to solutions which break the spherical symmetry near the droplet centre. We observe that random noise breaks the perfect spherical symmetry of the radial hedgehog solution near the centre, for small droplets and the predictions for large droplets are not much affected by random noise. 

The paper is organised as follows. In \Cref{sec:model}, we introduce the LdG framework studied in this paper. In \Cref{sec:nematic_square}, we study the problem of nematics within a square domain subject to tangent boundary conditions, in deterministic and stochastic settings. In \Cref{sec:deterministic}, we first summarise known results for the deterministic problem to give context to our stochastic results. In \Cref{sec:stochastic_setup}, the stochastic problem is introduced and we study the effects of additive and multiplicative noise. Using our stochastic framework, in \Cref{sec:switching}, we propose a method to model switching processes between the WORS, diagonal and rotated solutions. Finally, in \Cref{sec:droplet}, we introduce additive noise to a nematic-filled spherical droplet and look at its impact on the profile of the 
well- known radial hedgehog solution. In \Cref{sec:conclusions}, some conclusions and future directions are discussed for the reader's interest.

\section{Modelling framework }\label{sec:model}
In this paper we work in the LdG-framework, wherein the nematic state is described by the LdG $\Qvec$-tensor - a symmetric traceless $3\times 3$ matrix with five degrees of freedom, i.e., $\Qvec\in S_0:=\{\Qvec\in\mathbb{M}^{3\times3}:Q_{ij}=Q_{ji},  \sum_{i=1}^3 Q_{ii}=0\}$ (where $\mathbb{M}^{3\times3}$ denotes the space of all $3\times3$ matrices). Consequently, using the spectral decomposition theorem, $\Qvec$ can be expressed in terms of an orthonormal set of eigenvectors $\mathbf{n}_i$ and associated eigenvalues ${\lambda}_i$ as
\begin{equation}
    \Qvec=\lambda_1\mathbf{n}_1\otimes\mathbf{n}_1+\lambda_2\mathbf{n}_2\otimes\mathbf{n}_2+\lambda_3\mathbf{n}_3\otimes\mathbf{n}_3.
\end{equation}
Here, $(\nvec\otimes\nvec)_{ij}=n_i n_j$ is the vector tensor product. 
Similarly, $\mathbf{I}_3$ can be expressed as $\mathbf{I}_3=\sum^3_{i=1}\mathbf{n_i}\otimes\mathbf{n}_i$. Combining this with the constraint $\lambda_1+\lambda_2+\lambda_3=0$, we see
\begin{equation*}
    \Qvec=(2\lambda_1+\lambda_2)(\nvec_1\otimes\nvec_1)+(2\lambda_2+\lambda_1)(\nvec_2\otimes\nvec_2)-(\lambda_1+\lambda_2)\mathbf{I}_3.
\end{equation*}
Hence, setting 
\begin{subequations}\label{eq:order_paramters}
    \begin{align}
    s&=\lambda_1 -\lambda_3=2\lambda_1+\lambda_2, \\
    r&=\lambda_2 -\lambda_3=\lambda_1+2\lambda_2,
    \end{align}
\end{subequations}
$\Qvec$ can be written as
\begin{equation}
    \Qvec=s\left(\nvec_1\otimes\nvec_1-\frac{1}{3}\I_3\right)+r\left(\nvec_2\otimes\nvec_2-\frac{1}{3}\I_3\right).\label{eq:biaxial_Q}
\end{equation}
A biaxial phase is represented by a LdG $\Qvec$-tensor with three distinct eigenvalues and the director is defined to be the eigenvector of $\Qvec$ with the largest positive eigenvalue. The secondary director is modelled by the eigenvector with the second largest eigenvalue. 
A uniaxial nematic phase is modelled by a $\Qvec$-tensor with two equal non-zero eigenvalues and there is a single distinguished material direction, modelled by the uniaxial director which is the eigenvector with the non-degenerate eigenvalue. 
For example, if $\lambda_2=\lambda_3 \neq 0$, $\Qvec$ can be written as 
\begin{equation}
    \Qvec=s\left(\nvec_1\otimes\nvec_1-\frac{1}{3}\I_3\right),\label{eq:general_uniaxial_Q}
\end{equation}
where $\nvec$ is the uniaxial director or eigenvector with the non-degenerate eigenvalue, $\lambda_1$. 
Finally, the nematic phase is \emph{isotropic} if the $\Qvec$-tensor has three equal eigenvalues, so that $s=r=0$ and $\Qvec=\mathbf{0}$, for which all directions are physically equivalent \cite{majumdar-2010-article}. 

We measure the degree of biaxiality with the biaxiality parameter $\beta$, defined to be \cite{zarnescu-2010-article}
\begin{equation}\label{eq:beta}
    \beta:=1-6\frac{(\textrm{tr}\Qvec^3)^2}{(\textrm{tr}\Qvec^2)^3},
\end{equation} for $\Qvec\in S_0\setminus \{\mathbf{0}\}$.
From Lemma 1 in \cite{zarnescu-2010-article}, $\beta\in[0,1]$, $\beta=0$ for uniaxial and isotropic $\Qvec$-tensors, whereas non-zero $\beta$ is a signature of biaxiality. Further, $\beta$ attains its maximum value of unity when the $\Qvec$-tensor has a zero eigenvalue.

\section{Nematic liquid crystals confined to square domains}\label{sec:nematic_square}

In this section, we consider nematic liquid crystals confined to a 2D square domain, $\tilde{\Omega}=[-L,L]\times[-L,L]$ ($2L$ is the physical length of the square edges) in the $xy$-plane. This is appropriate for modelling three-dimensional square wells, for which the height of the well is much smaller than the square cross-sectional dimensions, and it is reasonable to assume that the structural details are invariant along the height of the well. 
Imposing planar surface anchoring conditions on the top and bottom of the well, which enforce tangent boundary conditions, along with $z$-invariant uniaxial Dirichlet boundary conditions on the lateral surfaces, it can be shown the space of physically relevant $\Qvec$-tensors is constrained to have the director lie in the $xy$-plane, and have $\mathbf{e}_z$ as a fixed eigenvector with associated fixed eigenvalue, in the limit of vanishing cell thickness \cite{golovaty2015}. With a fixed eigenvector, $\Qvec$ only has three degrees of freedom, $q_1,q_2,q_3$, and can be expressed as
\begin{equation} \label{eq:Q}
    \Qvec(\mathbf{x})=
    \begin{pmatrix}
    q_1-q_3  & q_2 & 0 \\
    q_2 & -q_1 -q_3  & 0 \\
    0 & 0 & 2q_3
    \end{pmatrix},\textrm{ for $\mathbf{x}\in\tilde{\Omega}$}.
\end{equation}
For formal arguments regarding the reduction from a 3D to a 2D problem for thin systems, and from five to three degrees of freedom, see \cite{golovaty2015,wang_canevari_majumdar_2019} and in particular, Theorem 5.1 and Theorem 2.1 respectively.

We work with a simple form of the LdG energy $F$, for which the energy density is the sum of a Dirichlet elastic energy density (to penalise spatial inhomogeneities) and a fourth order thermotropic bulk potential, $f_b$ \cite{canevari_majumdar_spicer2017,majumdar-2010-article}: 
\begin{equation}
    f_b(\Qvec)=\frac{A}{2}\textrm{tr}(\Qvec^2) -\frac{B}{3}\textrm{tr}(\Qvec^3) +\frac{C}{4}(\textrm{tr}\Qvec^2)^2,
\end{equation}
and
\begin{equation}\label{eq:dimensional_energy}
    F(\Qvec)=\int_{\tilde{\Omega}} \frac{K}{2}|\nabla \Qvec|^2 + f_b(\Qvec)~\mathrm{d}A,
\end{equation} where $|\nabla \Qvec|^2$ is the sum of the squares of the spatial derivatives of the components of the LdG $\Qvec$-tensor. 
Here, $K,B,C>0$ are material dependent constants independent of temperature, while $A$ depends linearly on temperature and is given by
\begin{equation*}
    A=\alpha(T-T^*),
\end{equation*}
where $\alpha>0$ is a material dependent constant, $T$ is the absolute temperature of the system, and $T^*$ is a characteristic liquid crystal temperature \cite{deGennes}. We work with $A<0$, i.e., low temperatures, so that an ordered uniaxial state is the globally stable critical point of $f_b$ \cite{mottram2014introduction}. In this case, the set of bulk energy minimisers is $\mathcal{N}=\{\Qvec\in S_0:\Qvec=s_+(\nvec\otimes\nvec-\I/3)\}$, with
\begin{equation}\label{eq:s_+}
    s_+=\frac{B+\sqrt{B^2-24AC}}{4C},
\end{equation}
and $\nvec\in\mathbb{S}^2$ (the unit sphere) is arbitrary \cite{majumdar-2010-article}. 

We non-dimensionalise the energy with the following change of variables $x=L\Bar{x}$, $y=L\Bar{y}$, and the dimensionless free energy is given by
\begin{equation}
    \bar{F}(\Qvec)=\frac{F(\Qvec)}{K}=\int_{\bar{\Omega}} \frac{1}{2}|\bar{\nabla} \Qvec|^2 + \frac{L^2}{K}f_b(\Qvec)~\mathrm{d}\bar{A},\label{eq:energy}
\end{equation}
with the re-scaled domain, $\bar{\Omega}=[-1,1]\times[-1,1]$.
Note that $\frac{[L^2]}{[K]}\times[A,B,C]=\frac{\textrm{m}^2}{\textrm{N}}$Nm$^{-2}$ is dimensionless. 
Henceforth, we drop bars and consider all quantities to be dimensionless. The Euler-Lagrange equations associated with \eqref{eq:energy} are:
\begin{subequations}\label{eq:EL-equations}
\begin{align}
    &\Laplace q_1=\frac{L^2}{K}\left(A q_1 +2B q_1 q_3 +C(2q_1^2+2q_2^2+6q_3^2)q_1\right), \label{eq:q_1} \\
    &\Laplace q_2=\frac{L^2}{K}\left(A q_2 +2B q_2 q_3 +C(2q_1^2+2q_2^2+6q_3^2)q_2\right), \label{eq:q_2} \\
    &\Laplace q_3=\frac{L^2}{K}\left(A q_3+B\left(\frac{1}{3}\left(q_1^2+q_2^2\right)-q_3^2\right) +C(2q_1^2+2q_2^2+6q_3^2)q_3\right) \label{eq:q_3} .
\end{align}
\end{subequations}
Using standard arguments in elliptic regularity, one can deduce that solutions of \eqref{eq:EL-equations} are real analytic in $\Omega$ \cite{zarnescu-2010-article}.



\begin{figure}[ht]
    \centering
    \begin{minipage}{0.35\textwidth}
        \centering
        \includegraphics[width=1.0\textwidth]{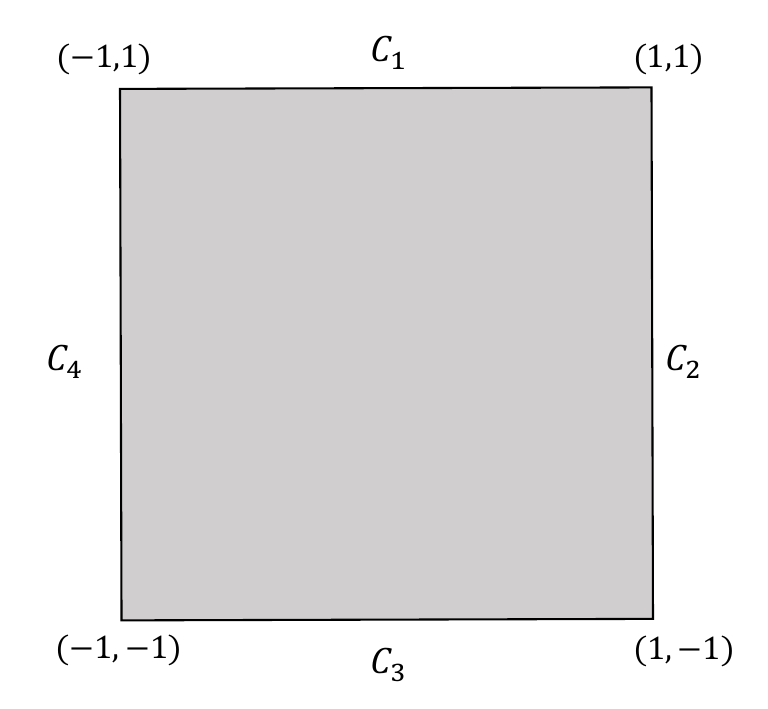}\\
    \end{minipage}
    \caption{Square geometry.}
    \label{fig:geometry}
\end{figure}

The square edges are labelled by $C_i$ for $i=1,2,3,4$, as in \Cref{fig:geometry}, i.e., 
\begin{align*}
    &C_1:=\{(x,1)\in\mathbb{R}^2:x\in(-1,1)\}, \\
    &C_2:=\{(1,y)\in\mathbb{R}^2:y\in(-1,1)\}, \\
    &C_3:=\{(x,-1)\in\mathbb{R}^2:x\in(-1,1)\}, \\
    &C_4:=\{(-1,y)\in\mathbb{R}^2:y\in(-1,1)\}, 
\end{align*}
and the set of corners/vertices is denoted by $E:=\{(1,1),(1,-1),(-1,-1),(-1,1)\}$. 
Following the existing literature \cite{tsakonas_brown_davidson_mottram, luo-2012, OR_kralj_majumdar}, we impose tangent uniaxial Dirichlet boundary conditions on the square edges
\begin{equation}
    \Qvec_\textbf{b}(\mathbf{x})=\begin{cases}
        s_+\left(\nvec_1\otimes\nvec_1-\frac{1}{3}\I_3\right) \quad\textrm{for $\textbf{x}\in C_1\cup C_3$}\\
        s_+\left(\nvec_2\otimes\nvec_2-\frac{1}{3}\I_3\right) \quad\textrm{for $\textbf{x}\in C_2\cup C_4$},
    \end{cases}
\end{equation}
where 
\begin{equation}
    \nvec_1=(1,0,0) \textrm{ and } \nvec_2=(0,1,0).
\end{equation}
At the square vertices, we set
\begin{equation}
    \Qvec_\mathbf{b}=\begin{pmatrix}
    -s_+/6  & 0 & 0 \\
    0 &  -s_+/6  & 0 \\
    0 & 0 & s_+/3
    \end{pmatrix}
    \quad\textrm{for } 
    \mathbf{x}\in E,
\end{equation}
to eliminate discontinuities from the conflicting tangent boundary conditions.
These boundary conditions translate to the following conditions on the components of $\Qvec$:
\begin{equation}
    q_1(x,y)=
    \begin{cases}
    \frac{s_+}{2} \textrm{ on } C_1\cup C_3,\\
    -\frac{s_+}{2} \textrm{ on } C_2\cup C_4,\\
    0\textrm{ on } E,
    \end{cases}
\end{equation}
$q_2=0$ on $\partial\Omega$ and $q_3=-\frac{s_+}{6}$ on $\partial\Omega$. 

For the remainder of the manuscript, we work with the special temperature, $A=\frac{-B^2}{3C}$, for which $s_+ = \frac{B}{C}$ and there is a solution branch of \eqref{eq:EL-equations} with constant $q_3=-\frac{s_+}{6}$, which is compatible with the boundary conditions for $q_3$. In this case, we numerically compute solution branches of \eqref{eq:EL-equations} i.e., $(q_1,q_2,q_3)=\left(q_1,q_2,-\frac{B}{6C}\right)$, by solving the gradient-flow equations
\begin{subequations}\label{eq:full_flow}
\begin{align}
    &\frac{\partial q_1}{\partial t}=\Laplace q_1-\frac{L^2 C}{K}\left(-\frac{2B^2}{3C^2} q_1  +\left(2q_1^2+2q_2^2+\frac{B^2}{6C^2}\right)q_1\right), \label{eq:q_1_flow} \\
    &\frac{\partial q_2}{\partial t}=\Laplace q_2-\frac{L^2 C}{K}\left(-\frac{2B^2}{3C^2} q_2 + \left(2q_1^2+2q_2^2+\frac{B^2}{6C^2}\right)q_2\right). \label{eq:q_2_flow}
\end{align}
\end{subequations}
with appropriate initial conditions, subject to the boundary conditions specified above. 
The principle here, is that for long enough times, solutions of the gradient flow model evolve to energy minimisers (or critical points) of the free energy \eqref{eq:energy}, i.e., they are steady solutions which satisfy $\frac{\partial q_i}{\partial t}=0$ for $i=1,2$, so that $\left(q_1,q_2,-\frac{B}{6C}\right)$ is a solution of \eqref{eq:EL-equations}. Henceforth, we define $\tilde{L}=\frac{2L^2C}{K}$ to be the single dimensionless parameter, which is interpreted as a measure of domain size due to its dependence on $L$.
Solving \eqref{eq:full_flow} numerically, we recover the diagonal and rotated solutions first reported in \cite{tsakonas_brown_davidson_mottram}, and later recovered in \cite{luo-2012}, and both solutions have non-zero $q_2$, by using different initial conditions.

Additionally, there is also a specific solution branch with $q_2=0$, which satisfies \eqref{eq:q_2} and is compatible with the boundary conditions. Hence, there also exists a solution branch of the form $(q_1,q_2,q_3)=\left(q,0,-\frac{B}{6C}\right)$ for all $\tilde{L}\geq0$. The single governing equation for $q$ is, thus,
\begin{equation}
    \Laplace q =\frac{L^2}{K}\left(2C q^3 -\frac{B^2}{2C}q\right). \label{eq:q}
\end{equation}
This solution branch includes the well order reconstruction solution (WORS) reported in \cite{OR_kralj_majumdar, canevari_majumdar_spicer2017}, with $q=0$ along the square diagonals so that the WORS has a uniaxial diagonal cross (see \eqref{eq:Q} with $q_1 = q_2 = 0$ along the square diagonals to see why there is a uniaxial diagonal cross for the WORS) and this is surrounded by regions of maximal biaxiality.
To find solutions of \eqref{eq:q}, we solve the corresponding gradient flow equation:
\begin{equation}
    \frac{\partial q}{\partial t}=\Laplace q-\frac{2L^2 C}{K} q \left(q-\frac{B}{2C}\right)\left(q+\frac{B}{2C}\right),\label{eq:q_flow}
\end{equation}
i.e., the Allen-Cahn equation. Note, any steady solution $q$ of \eqref{eq:q_flow}, is also a steady solution of \eqref{eq:full_flow}, with $q_1=q$ and $q_2=0$.


\subsection{Deterministic case}\label{sec:deterministic}

\subsubsection{Numerical method}\label{sec:deterministic_method}
We solve the gradient flow equations \eqref{eq:full_flow} in Matlab using finite difference methods and the Runge-Kutta method \cite{RK_reference}. We divide $\Omega=[-1,1]\times[-1,1]$ into a uniform square grid of points, where 
$k=1/(N+1)$ is the spatial step size. Our time interval is $[0,T]$, $N_t$ is the total number of time steps and $\Delta t=T/N_t$ is the time step size.  Considering \eqref{eq:q_flow} for example, the approximation of the solution $q$, at a point $(x_{j_1}, y_{j_2})\in(-1,1)$, at time $t_n=n\Delta t$, is denoted by $q^{j_1,j_2}_n$. $q_n$ is a $2N+1\times 2N+1$ matrix, whose $j_1^{th},j_2^{th}$ element is $q^{j_1,j_2}_n$. Here, $j_1 ,j_2 =1,\ldots,2N+1$ and $n=1,\ldots,N_t$.
The Runge-Kutta method is then defined by
\begin{equation}
    q_{n+1}=q_{n}+\frac{\Delta t}{6}(k_1+2k_2+2k_3 +k_3),
\end{equation}
where
\begin{subequations}\label{eq:k_i}
\begin{align}
    & k_1=g(t_n,q_n),\\
    & k_2=g(t_n+\Delta t/2,q_{n}+\Delta t k_1/2),\\
    & k_3=g(t_n+\Delta t/2,q_{n}+\Delta t k_2/2),\\
    & k_4=g(t_{n+1},q_{n}+\Delta t k_3), 
\end{align}
\end{subequations}
and (using Matlab notation)
\begin{align}
    g(t_n,q_n)&=\Bigg(\left[-\frac{B}{2C}\textrm{ones}(2N+1,1),q_n(:,1:2N)\right] \nonumber \\
    & \qquad + \left[q_n(:,2: 2N+1),-\frac{B}{2C} \textrm{ones}(2N+1,1)\right] \nonumber \\
    & \qquad + \left[\frac{B}{2C}\textrm{ones}(1,2N+1) ; q_n(1:2N,:) \right] \nonumber \\& \qquad + \left[ q_n(2: 2N+1,:) ; \frac{B}{2C}\textrm{ones}(1,2N+1) \right] -4q_n \Bigg)/k^2\label{eq:g}
    -f(q_{n}),
\end{align}
for $f(q)=\tilde{L} q \left(q-\frac{B}{2C}\right)\left(q+\frac{B}{2C}\right)$. The numerical method for solving the system \eqref{eq:full_flow} is analogous, with modifications to $g$ and $f$ to account for the different right hand sides in \eqref{eq:q_1_flow} and \eqref{eq:q_2_flow}.

\subsubsection{Summary of the deterministic solution landscape}\label{sec:deterministic_numerics}

To put our stochastic results into context, we first summarise the solution landscape of \eqref{eq:EL-equations} (see \cite{canevari_majumdar_spicer2017, tsakonas_brown_davidson_mottram, 2D_landscape} for the original work). We numerically compute the solution branches $(q_1,q_2,q_3)=\left(q,0,-\frac{B}{6C}\right)$ and $(q_1,q_2,q_3)=\left(q_1,q_2,-\frac{B}{6C}\right)$. The numerical solution of \eqref{eq:q_flow} or \eqref{eq:full_flow}, is deemed to be steady, or equivalently a solution of \eqref{eq:EL-equations}, if the norm of the gradient falls below $10^{-6}$, i.e., the norm of the right hand side of \eqref{eq:q_flow} or \eqref{eq:full_flow} falls below $10^{-6}$. At this point, we assume that our numerical method has converged. We take $N=79$ and $\Delta t=2\times 10^{-5}$. With appropriate initial conditions, we observe that our numerical solutions converge for $T \leq 2$. Throughout this section and \Cref{sec:stochastic_setup}, $A=-\frac{B^2}{3C}$, $B=0.64\times10^4$N, $C=0.35\times10^4$N \cite{majumdar-2010-article, canevari_majumdar_spicer2017} and $K=10^{-11}$N \cite{zarnescu-2010-article}, so that varying $\tilde{L}$ is equivalent to changing the square domain size. When $\tilde{L}=0.05$, the corresponding square edge length is $8.45\times 10^{-9}$m yielding a nano-scale square domain,  whilst for $\tilde{L}=200$, the corresponding edge length is $5.35\times 10^{-7}$m and the square is closer to an experimentally achievable micron-scale square domain. 
Here and in our stochastic results, we plot the biaxiality parameter $\beta$ defined in \eqref{eq:beta} and the director $\nvec$, taken to be the eigenvector of $\Qvec$ with largest positive eigenvalue. Consequently, the director must lie in the $xy$-plane (since $\Qvec$ is negatively ordered in the $\mathbf{e}_z$ direction with eigenvalue $2q_3=-B/6C$, throughout $\Omega$) and can therefore be defined as $\nvec=(\cos\theta,\sin\theta)$, where 
$\theta=\textrm{atan2}(q_2,q_1)$ is the angle between the director and the $x$-axis.

\begin{figure}
    \centering
        \includegraphics[width=1.0\textwidth]{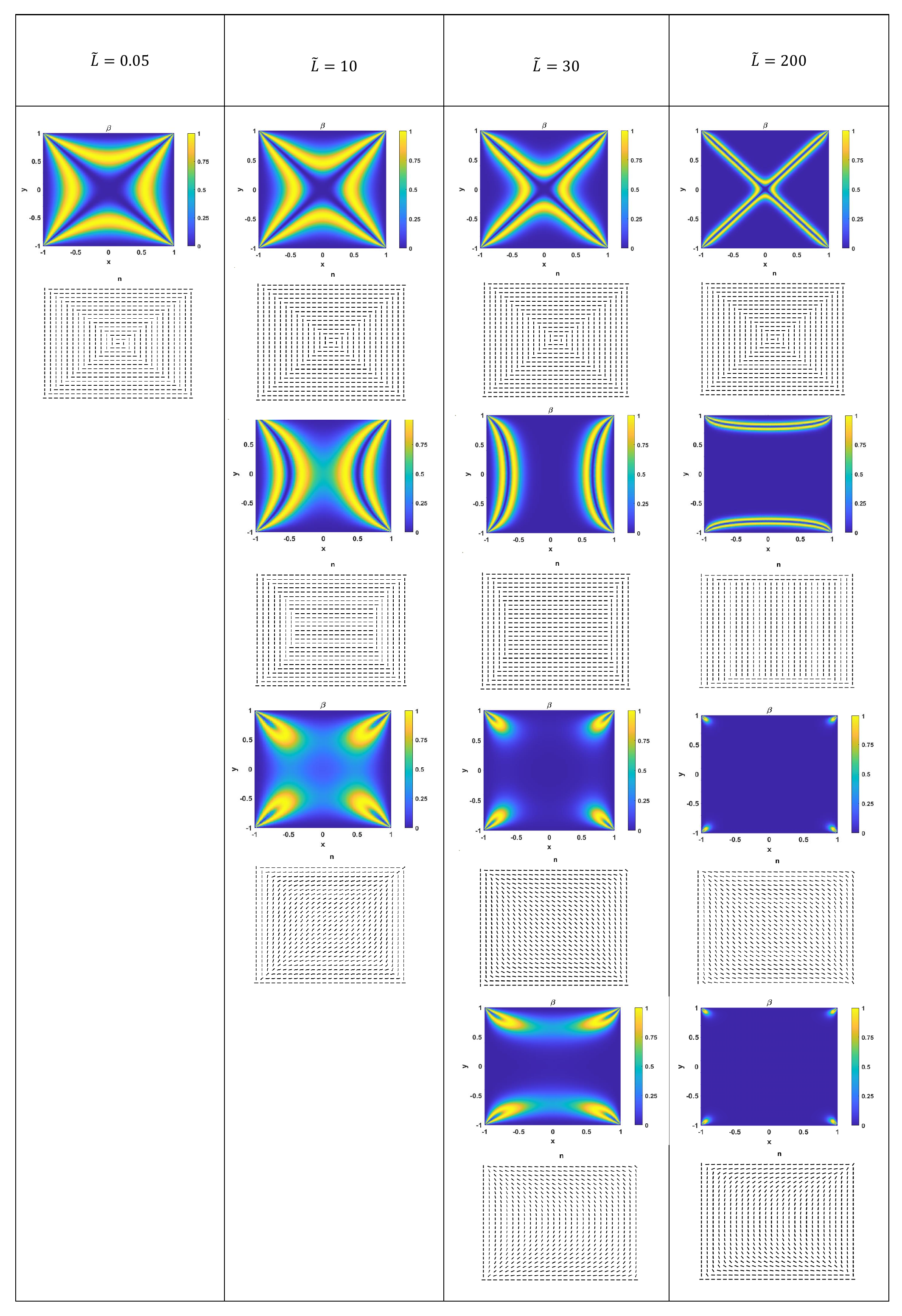}
        \caption{Summary of the solutions to \eqref{eq:EL-equations} (the deterministic problem) for fixed $q_3=-B/6C$. Here, we plot the biaxiality parameter, $\beta$, and director, $\nvec$, of solutions for the stated values of $\Tilde{L}$. Top row: WORS solutions, second row from left to right: BD$_y$, BD$_y$, BD$_x$, third row: diagonal solutions, fourth row: rotated solutions.}
    \label{fig:summary}
\end{figure}

For solution branches with $q_2 = 0$ i.e. $(q_1,q_2,q_3)=\left(q,0,-\frac{B}{6C}\right)$, we find the WORS and bent-director (BD) solutions. With $q_2 = 0$, the corresponding LdG $\Qvec$-tensor has three constant eigenvectors along the coordinate directions. The defining feature of the WORS is two lines of uniaxiality along the square diagonals, 
and this is surrounded by regions of biaxiality including maximal biaxiality (see \Cref{fig:summary}, first row). On the square diagonals, $q_1:=q=0$ and $q_2=0$ so that
\begin{equation}
    \Qvec=
    \begin{pmatrix}
    \frac{B}{6C}  & 0 & 0 \\
    0 &  \frac{B}{6C}  & 0 \\
    0 & 0 & -\frac{B}{3C}
    \end{pmatrix},
\end{equation}
i.e. $\Qvec$ is uniaxial on the square diagonals, for the WORS.
Numerically, we classify a solution as being WORS if the average value of $|q|$ on each of the square diagonals, and the average value of $|q_2|$ throughout $\Omega$ (i.e., the sum of the absolute value of the numerical solution $q$ ($q_2$) at every point on the square diagonal (in $\Omega$) divided by the total number of points on the square diagonal (in $\Omega$)), are both less than $10^{-6}$. This is an arbitrary measure that suffices for our numerical experiments.
There are two types of BD solutions, with two bands of biaxiality near a pair of parallel edges. 
We label the BD solution in \Cref{fig:summary} (second row) for large $\tilde{L}=200$, as BD$_x$, since the bands of biaxiality are parallel to the $x$-axis. Similarly, we label the BD solutions in \Cref{fig:summary} for intermediate $\tilde{L}=10,30$, as BD$_y$. 
The solution branches with non-zero $q_2$, denoted by $(q_1,q_2,q_3)=\left(q_1,q_2,-\frac{B}{6C}\right)$, include the diagonal and rotated solutions. There are two diagonal solutions for which the director aligns along one of the square diagonals (see \Cref{fig:summary}, third row) and four rotated solutions (see \Cref{fig:summary}, fourth row) where the director rotates by $\pi$ radians between a pair of parallel edges (we only present two of the four rotated solutions). The diagonal and rotated solutions exhibit biaxiality near the square vertices and the biaxial regions shrink as $\tilde{L}$ increases (this can be seen looking across the rows in \Cref{fig:summary}).

\begin{figure}[ht]
    \centering
    \begin{minipage}{1.0\textwidth}
        \centering
        \includegraphics[width=0.45\textwidth]{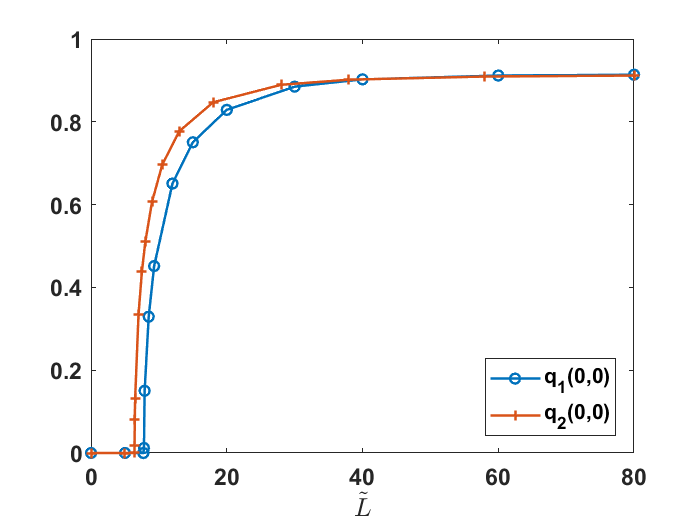}\\
    \end{minipage}
    \caption{Plot of $q_1(0,0)$ and $q_2(0,0)$ as a function of $\tilde{L}$. $q_2(0,0)$ becomes non-zero (indicating $q_2$ is non-zero in $\Omega$) for $\tilde{L}\approx 6.4$ indicating the WORS is unstable beyond this point due to the emergence of diagonal solutions. $q_1(0,0)$ becomes non-zero for $\tilde{L}\approx 7.7$, indicating the emergence of BD solutions.} 
    \label{fig:bifurcation_diagram}
\end{figure}

The WORS exists for all $\tilde{L}\geq0$, it is the unique and globally stable critical point of \eqref{eq:energy} (i.e., all the eigenvalues of the Hessian of the free energy at that critical point are positive, otherwise a critical point is unstable) for sufficiently small $\tilde{L}$, but loses stability for sufficiently large $\tilde{L}$ \cite{canevari_majumdar_spicer2017}. In \Cref{fig:bifurcation_diagram}, we track the value of 
$q_1(0,0)$ and $q_2(0,0)$ as a function of $\tilde{L}$. To compute this plot, we use an initial condition which favours $q_1(0,0)\neq 0$ (\eqref{eq:WORS_IC} with $q^0_1=0.1$ along the diagonals) and $q_2(0,0)\neq 0$ (\eqref{eq:WORS_IC} and $q_2^0=0.9$). 
Hence, the value of $\tilde{L}$ for which 
$q_1(0,0)\neq0$ or $q_2(0,0)\neq0$ (whichever occurs first) labels the bifurcation point at which the WORS loses stability and new solutions emerge. This critical value is numerically computed to be $\tilde{L}_c\approx 6.4$. For $\tilde{L}_c > 6.4$, $q_2(0,0)\neq0$ due to the emergence of diagonal solutions. For $\tilde{L}\geq 7.8$, BD solutions emerge, followed by the appearance of rotated solutions for $\tilde{L}\geq 28$. 

With the following initial condition for \eqref{eq:q_flow},
\begin{equation}\label{eq:WORS_IC}
    q^0(x,y)=\begin{cases}
    \frac{B}{2C}\;&\textrm{ for }-|y|<x<|y|\\
    -\frac{B}{2C}\;&\textrm{  for }-|x|<y<|x|\\
    0 \;&\textrm{     for }|y|=|x|,
    \end{cases}
\end{equation}
we can numerically compute the WORS with $q=0$ on the square diagonals and $q_2 = 0$ throughout the domain, for all $t$ and for all 
 positive $\tilde{L}$. To find BD solutions, we take $q^0=\pm0.9$ on $|y|=|x|$ in \eqref{eq:WORS_IC} and solve the gradient-flow model \eqref{eq:q_flow}. In the context of the full system \eqref{eq:full_flow}, we additionally take $q^0_2=0$ everywhere as our initial condition to find WORS and BD solutions. 
To find diagonal solutions of \eqref{eq:full_flow}, we take \eqref{eq:WORS_IC} (for $q^0_1$) and $q^0_2=\pm 0.9$ at every mesh point as our initial condition. To find rotated solutions of \eqref{eq:full_flow}, we  solve $\Laplace \theta_0=0$ 
subject to  
\begin{subequations}\label{eq:theta_BCs}
\begin{align}
    &\theta_0(x,1)=\pi,\; \theta_0(x,-1)=0 \textrm{ for } -1< x<1 \\
    &\theta_0(1,y)=\theta_0(-1,y)=\frac{\pi}{2} \textrm{ for } -1< y<1,\\
    &\theta_0(\pm 1,\pm 1)=0.
\end{align}
\end{subequations}
Setting $\nvec_0=(\cos\theta_0,\sin\theta_0,0)$ and $\Qvec^0=s_+\left(\nvec_0\otimes\nvec_0-\frac{1}{3}\I\right)$, our initial conditions are $q^0_1=Q^0_{11}$ and $q^0_2=Q^0_{12}$. By altering the boundary conditions in \eqref{eq:theta_BCs} accordingly, we find the remaining three rotated solutions. Henceforth, we refer to these initial conditions as a WORS initial condition, BD initial condition, diagonal initial condition and rotated initial condition respectively, for brevity. 

\subsection{Stochastic case}\label{sec:stochastic_setup}
Next, we consider the effects of random noise on the solution branches discussed in the preceding section. 
A stochastic version of \eqref{eq:q_flow} is given by
\begin{equation}\label{eq:stochastic_q}
    d q=\left[\Laplace q-\tilde{L} q \left(q-\frac{B}{2C}\right)\left(q+\frac{B}{2C}\right)\right]dt+\sigma G(q) dW(t,\mathbf{x}),\quad t>0,\; \mathbf{x}\in\Omega
\end{equation}
where $W(t,\mathbf{x})$ is known as a $Q$-Wiener process (which introduces random noise), $\sigma$ controls the strength of the noise and $G$ is a function which may depend on the solution $q$ \cite{into_book}. Informally speaking, a $Q$-Wiener process is an example of coloured noise which has correlation in space, as opposed to white noise which is homogeneous in space. We study a $Q$-Wiener process as opposed to white-noise because, the two-dimensional problem with white noise is ill-posed and further treatment is needed (see \cite{hairer2012triviality,ryser2012well} for instance). 
The physical interpretation of the $Q$-Wiener process $W(t,\mathbf{x})$, is that it is a random term in both space and time, 
which accounts for random fluctuations found in nature \cite{into_book}. Therefore, in our context, we can interpret it as uncertainties in material/system properties. 
For a formal definition of a $Q$-Wiener process see \cite{into_book}. Similarly, a stochastic version of \eqref{eq:full_flow} is 
\begin{subequations}\label{eq:full_stochastic}
\begin{align}
    &dq_1=\left[\Laplace q_1-\frac{\tilde{L}}{2}\left(-\frac{2B^2}{3C^2} q_1  +\left(2q_1^2+2q_2^2+\frac{B^2}{6C^2}\right)q_1\right)\right]dt + \sigma G_1(q_1,q_2) d W, \label{eq:q_1_stochastic} \\
    &d q_2=\left[\Laplace q_2-\frac{\tilde{L}}{2}\left(-\frac{2B^2}{3C^2} q_2 + \left(2q_1^2+2q_2^2+\frac{B^2}{6C^2}\right)q_2\right)\right]dt + \sigma G_2(q_1,q_2)d W. \label{eq:q_2_stochastic}
\end{align}
\end{subequations}

\subsubsection{Numerical method}

 

Following Theorem 10.7 in \cite{into_book}, a $Q$-Wiener process $W(t,\mathbf{x})$ can be expressed by the following sum
\begin{equation}\label{eq:Q-Wiener_process}
    W(t,\mathbf{x}):=\sum_{j=1}^\infty \sqrt{p_j}\chi_j(\mathbf{x})B_j(t).
\end{equation}
Here, $p_j$ and $\chi_j$ are the eigenvalues and eigenfunctions of the $Q$-Wiener process respectively, while $B_j$ are independent identically distributed (iid) Brownian motions (informally, a Brownian motion is a stochastic process whose increments are independent and normal distributed, see \cite{into_book} for a full definition).
We numerically implement the $Q$-Wiener process \eqref{eq:Q-Wiener_process} by a finite sum approximation 
as in Example 10.12 of \cite{into_book}. 

For $\Omega=[-1,1]\times[-1,1]$ and $U=L^2(\Omega)$, let $Q:U\to U$ be a bounded linear operator with eigenfunctions $\chi_{m_1,m_2}(\mathbf{x})= \frac{1}{2}e^{\pi i m_1 x}e^{\pi i m_2 y}$ and eigenvalues $p_{m_1,m_2}=e^{-\alpha \gamma_{m_1,m_2}}$, for a parameter $\alpha>0$ which controls the rate of decay of the noise and $\gamma_{m_1,m_2}=m_1^2+m_2^2$. 
We then let
\begin{equation}
    W^{N+1}(t,\mathbf{x}):=\sum_{m_1=-N+1}^{N+1}\sum_{m_2=-N+1}^{N+1}\sqrt{p_{m_1,m_2}}\chi_{m_1,m_2}(\mathbf{x})B_{m_1,m_2}(t),
\end{equation}
for iid Brownian motions $B_{m_1,m_2}(t)$, be our finite sum approximation. The difference $\Delta W^{N+1}_n=W^{N+1}(t+\Delta t, (x_{j_1},y_{j_2}))-W^{N+1}(t, (x_{j_1},y_{j_2}))$ approximates $dW(t,(x_{j_1},y_{j_2}))$ and this is computed using Algorithms 10.5 and 10.6 in \cite{into_book}. Considering \eqref{eq:stochastic_q}, we then add this approximation of $dW$ to our Runge-Kutta method (explained in \Cref{sec:deterministic_method}) at each time step, i.e.,

\begin{equation}
    q_{n+1}=q_{n}+\frac{\Delta t}{6}(k_1+2k_2+2k_3 +k_4),
\end{equation}
where $k_1,k_2,k_3,k_4$ are as in \eqref{eq:k_i}, and we add
\begin{align*}
    \sigma \Delta W^{N+1}_n /\Delta t,
\end{align*}
to 
each of the $k_j$, for $j=1,2,3,4$, following \cite{gard1988introduction, kloeden1992stochastic}. The same modifications apply to \eqref{eq:full_stochastic}. Due to the presence of noise, the order of convergence cannot exceed one \cite{kloeden1992stochastic}. An important concept for ergodic theory in random dynamical systems is the invariant measure, which refers to the limiting distribution that the stochastic solution obeys in the long run \cite{arnold1995random}. In other words, the time-dependent distribution of the evolving stochastic solution becomes time-invariant as time approaches infinity. \cite[Proposition 4.1]{wu2018random} guarantees the existence of the invariant measure of an equation such as \eqref{eq:stochastic_q} by showing the existence of the stationary solution, which gives rise to the invariant measure. The stability of the empirical distributions through numerical simulations in Section \ref{sec:additive} (see Figure \ref{fig:empirical_density}) illustrates the convergence of the numerical scheme to the invariant measure.


\subsubsection{Effects of additive noise}\label{sec:additive}

We begin by numerically exploring solutions of the stochastic equations \eqref{eq:stochastic_q} and \eqref{eq:q_1_stochastic}, \eqref{eq:q_2_stochastic}, under the addition of additive noise i.e, when $G(q)=G_1(q_1,q_2)=G_2(q_1,q_2)=1$, and assess the impact on the corresponding solution landscape. 
Recall, $\alpha>0$ appears in the eigenvalues of the $Q$-Wiener process and subsequently controls the rate of decay and spatial variation of the noise as well as its strength, while $\sigma$ multiplies $dW$ and consequently scales the strength of the noise. Since both $\alpha$ and $\sigma$ measure the strength of the noise, for simplicity, we fix $\sigma=1$ and vary $\alpha$ to control the noise amplitude. Large values of $\alpha$ (say $\alpha> 1$) represent weak noise with little spatial variation and small values of $\alpha$ (say $\alpha\leq 0.1$) represent strong noise with large spatial variation. 

\begin{figure}
    \centering
    \begin{minipage}{0.4\textwidth}
        \centering
        \includegraphics[width=1.0\textwidth]{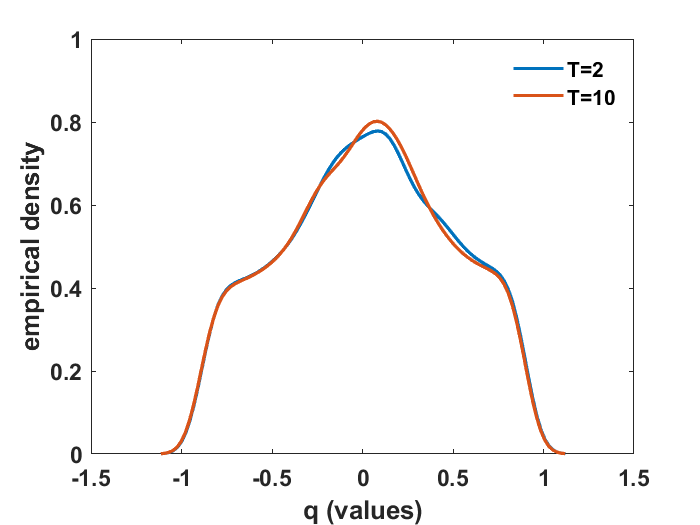}\\
    \end{minipage}
    \caption{Average of 100 empirical density functions at $T=2$ and $T=10$, for $\alpha=1$, $\sigma=1$ and $\tilde{L}=0.05$. } 
    \label{fig:empirical_density}
\end{figure}

In \Cref{sec:deterministic_numerics}, we consider steady solutions of the gradient flow equations \eqref{eq:q_flow} and \eqref{eq:full_flow}, which correspond to solutions of the Euler-Lagrange equations \eqref{eq:EL-equations}. For the stochastic case, as $dW$ introduces random fluctuations to the equations, the same notion of a steady solution does not apply. Instead, we consider solutions of \eqref{eq:stochastic_q} and \eqref{eq:full_stochastic} at a given time $T$. Hence, a solution of our stochastic partial differential equations is obtained by stopping our numerical method at a given time $T$.  Since the numerical solutions typically converge for $T\leq 2$ in the deterministic case, we take $\Delta t=2\times 10^{-5}$ again and plot solutions for the stochastic equations for $T=2$, so that these solutions are, in some sense, long-time equilibrium profiles. This claim is validated in \Cref{fig:empirical_density} (at least for certain parameter values), where the average of 100 empirical density functions is generated from 100 corresponding sample solutions of \eqref{eq:stochastic_q} (with $\alpha=1$, $\sigma=1$, $\tilde{L}=0.05$ and a WORS initial condition) 
at $T=2$ and $T=10$. The empirical density function is the probability density function associated to the numerical solution $q$, so that the area under the curve between two point $a$ and $b$, equals $\mathbb{P}(a\leq q\leq b)$ (i.e. the probability that the value of the numerical solution $q$, at any point in $\Omega$, is between $a$ and $b$), while the total area under the curve is 1. The curves at $T=2$ and $T=10$ are almost identical indicating the existence of an invariant measure \cite{Yue_invariant_measure}, meaning that over many simulations, there is little difference between solutions at $T=2$ and $T=10$.

\begin{figure}
\centering
    \begin{minipage}{0.3\textwidth}
        \centering
        \includegraphics[width=1.0\textwidth]{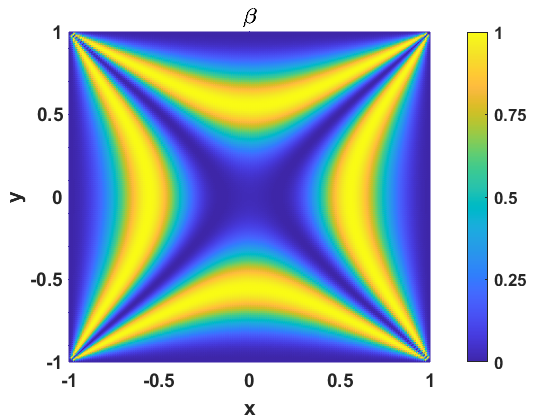}
    \end{minipage}
    \begin{minipage}{0.3\textwidth}
    \centering
        \includegraphics[width=1.0\textwidth]{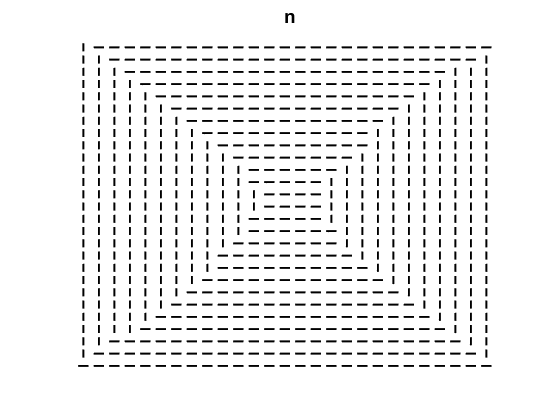}
    \end{minipage}\\
    \begin{minipage}{0.3\textwidth}
        \centering
        \includegraphics[width=1.0\textwidth]{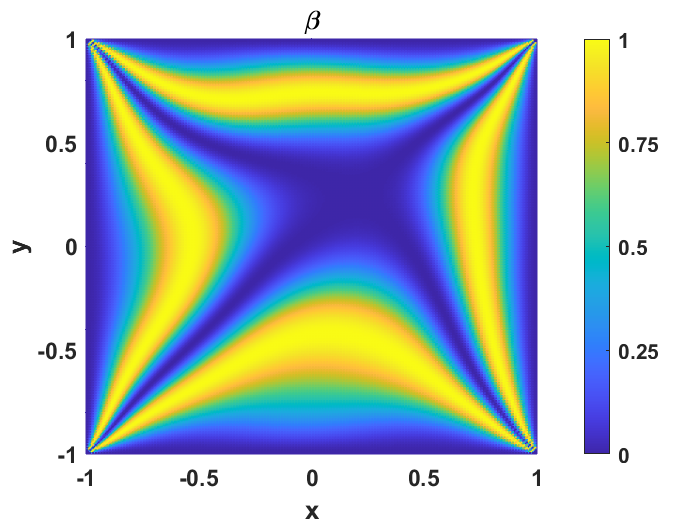}
    \end{minipage}
    \begin{minipage}{0.3\textwidth}
    \centering
        \includegraphics[width=1.0\textwidth]{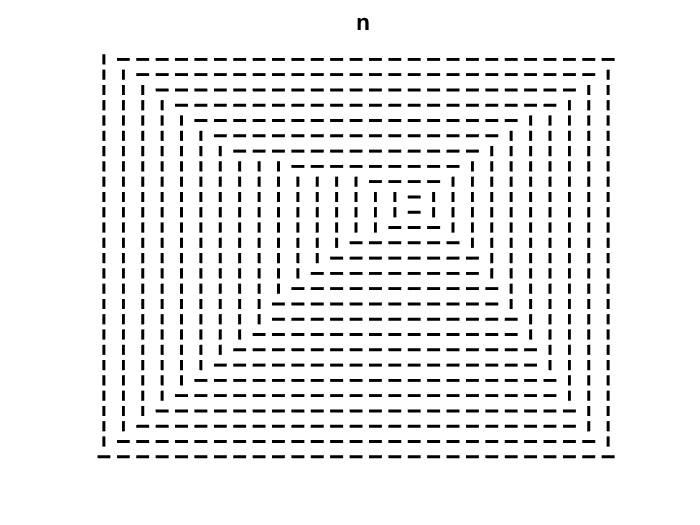}
    \end{minipage}
    \caption{First row: biaxiality parameter and director of an approximate WORS solution to \eqref{eq:stochastic_q} (under additive noise) for $\tilde{L}=0.05$, $\sigma=1$, $\alpha=3$ and $T=2$; $q(0,0)=0.0268$ for this solution. Second row: biaxiality parameter and director of an approximate WORS solution to \eqref{eq:stochastic_q} (under additive noise) for $\tilde{L}=0.05$, $\sigma=1$, $\alpha=0.1$ and $T=2$; $q(0,0)=0.0281$ for this solution.} 
    \label{fig:lambda_0.05_alpha_3_WORS}
\end{figure}

\begin{figure}
    \centering
    \begin{minipage}{0.3\textwidth}
        \centering
        \includegraphics[width=1.0\textwidth]{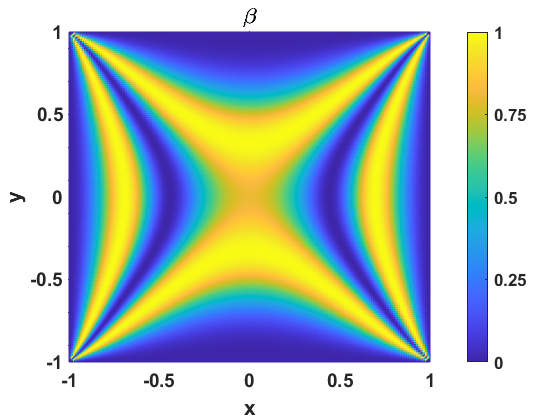}\\
    \end{minipage}
    \begin{minipage}{0.3\textwidth}
    \centering
        \includegraphics[width=1.0\textwidth]{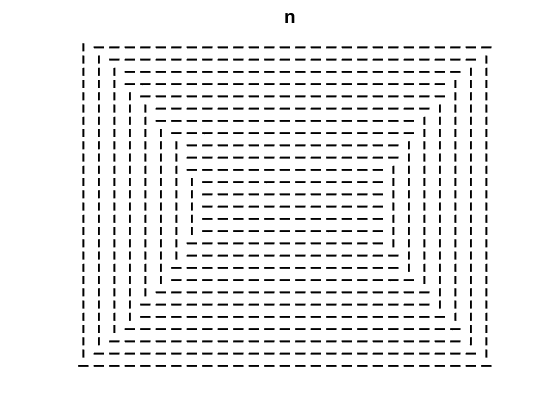}\\
    \end{minipage}\\
    \begin{minipage}{0.3\textwidth}
        \centering
        \includegraphics[width=1.0\textwidth]{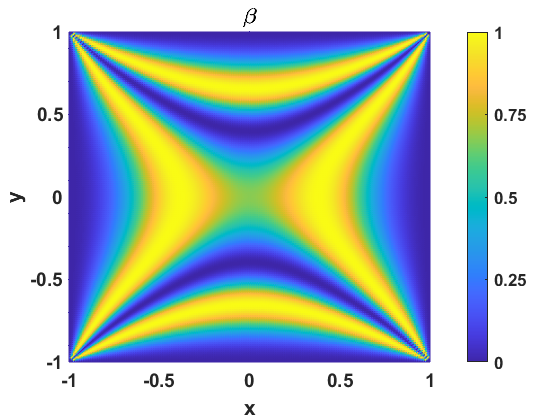}\\
    \end{minipage}
    \begin{minipage}{0.3\textwidth}
    \centering
        \includegraphics[width=1.0\textwidth]{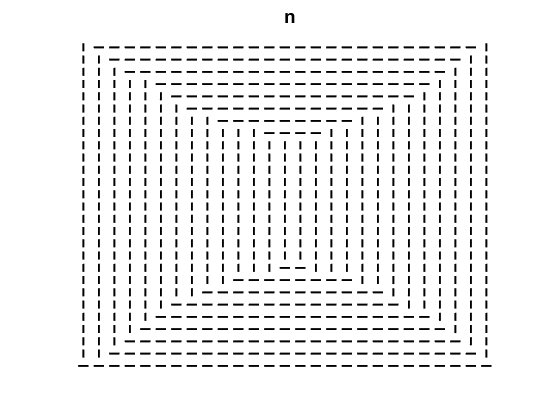}\\
    \end{minipage}
    \caption{First row: biaxiality parameter and director of a BD$_y$ solution to \eqref{eq:stochastic_q} (under additive noise) for $\tilde{L}=0.05$, $\sigma=1$, $\alpha=3$ and $T=2$. Second row: biaxiality parameter and director of a BD$_x$ solution to \eqref{eq:stochastic_q} (under additive noise) for $\tilde{L}=0.05$, $\sigma=1$, $\alpha=3$ and $T=2$.} 
    \label{fig:lambda_0.05_alpha_3_BD}
\end{figure}

We first study \eqref{eq:stochastic_q}, which is somewhat artificial as it assumes $q_2\equiv 0$ despite the inclusion of noise. This is different to studying \eqref{eq:full_stochastic}, since the noise reduces the probability of observing solutions with $q_2\equiv 0$.
The WORS is the unique stable solution of \eqref{eq:EL-equations} (with $q_3=-\frac{B}{6C}$) for small $\tilde{L}<6.4$. However, in the stochastic case, with $\tilde{L}=0.05$, weak additive noise $(\alpha=3)$ and a WORS initial condition, the WORS is no longer (qualitatively) unique as there are solutions of \eqref{eq:stochastic_q} which can be classified as either BD$_x$ or BD$_y$ (see \Cref{fig:lambda_0.05_alpha_3_BD}). In fact, as soon as noise is introduced, the uniaxial diagonal cross is lost (i.e, $q\neq0$ along the square diagonals) and one can only observe approximate WORS solutions. We classify a solution as being an approximate WORS if $|q(0,0)|<0.05$ and the average value of $|q|$ on each of the square diagonals is less than 0.05 (see \Cref{fig:lambda_0.05_alpha_3_WORS} first row). 
Decreasing $\alpha=0.1$ with $\tilde{L}=0.05$ (using a WORS inital condition still), the symmetry of the approximate WORS solution is further lost (\Cref{fig:lambda_0.05_alpha_3_WORS} second row), 
and finally with $\alpha=0.01$ (strong noise), the obtained solution is dominated by noise and completely random, hence approximate WORS solutions cannot be found. Similarly, for $\tilde{L}=0.05$ and $\alpha=0.1$ (using a BD initial condition), we see the symmetry of the biaxial bands along parallel square edges is lost for the BD solutions, whilst for $\alpha=0.01$, BD solutions cannot be found due to the large strength of the noise (these solutions are not presented). 

\begin{figure}[ht]
    \centering
    \begin{minipage}{0.32\textwidth}
        \centering
        \includegraphics[width=1.0\textwidth]{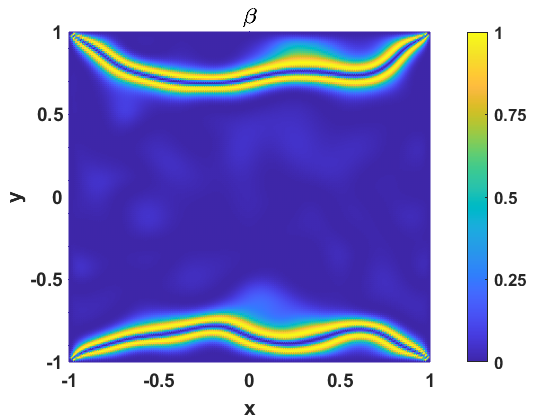}
    \end{minipage}
    \begin{minipage}{0.32\textwidth}
    \centering
        \includegraphics[width=1.0\textwidth]{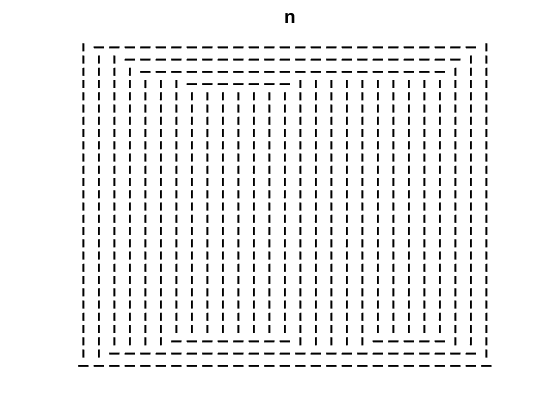}
    \end{minipage}
    \caption{Biaxiality parameter and director of a BD$_x$ solution to \eqref{eq:stochastic_q} (under additive noise) for $\tilde{L}=200$, $\sigma=1$, $\alpha=0.01$ and $T=2$.} 
    \label{fig:lambda_200_alpha_0.01}
\end{figure}

The existence of approximate WORS solutions also depends on $\tilde{L}$. In fact, we are unable to find such solutions for $\tilde{L}\geq 6.5$ with $\alpha=3$ (in 20 simulations with these parameter values and using a WORS inital condition, we did not observe an approximate WORS solution). This is again in contradiction to the deterministic picture where the WORS exists for all $\tilde{L}\geq 0$. BD solutions however, can be found for all values of $\tilde{L}$ in both the deterministic and stochastic settings (for suitably sized noise). Using a BD initial condition, for small $L=0.05$ and weak noise ($\alpha=3$), we observe BD solutions in \Cref{fig:lambda_0.05_alpha_3_BD}, and for large $\tilde{L}=200$ and strong noise ($\alpha=0.01$), we find solutions that are still clearly identifiable as BD (see \Cref{fig:lambda_200_alpha_0.01} for a BD$_x$ solution for instance). We speculate that for a fixed $\tilde{L}$, BD-solutions exist for \eqref{eq:stochastic_q} for $\alpha \geq \alpha_c(\tilde{L})$ where $\alpha_c$ is a decreasing function of $\tilde{L}$ i.e. the critical noise strength increases with increasing square edge length.

\begin{figure}
    \centering
    \begin{minipage}{0.3\textwidth}
        \centering
        \includegraphics[width=1.0\textwidth]{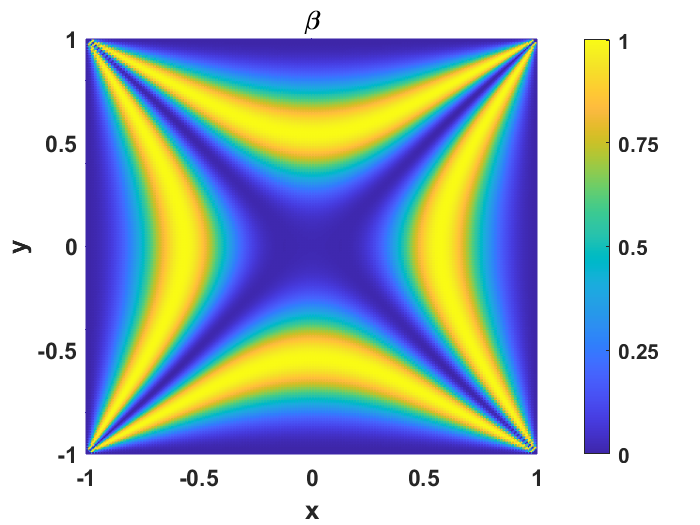}
    \end{minipage}
    \begin{minipage}{0.3\textwidth}
    \centering
        \includegraphics[width=1.0\textwidth]{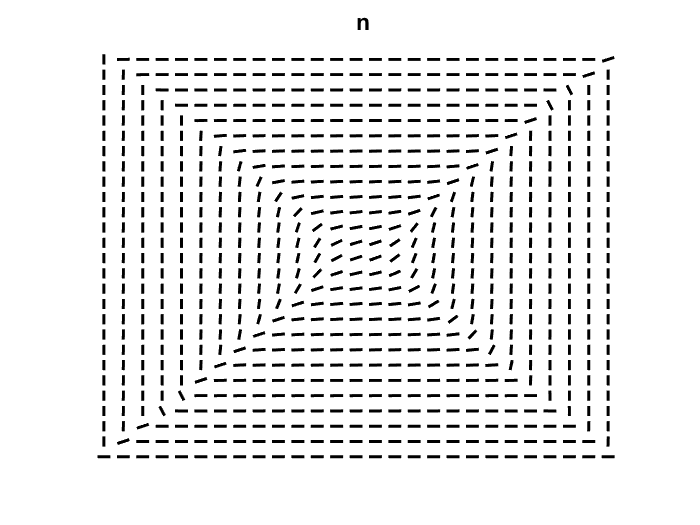}
    \end{minipage}\\
    \begin{minipage}{0.3\textwidth}
        \centering
        \includegraphics[width=1.0\textwidth]{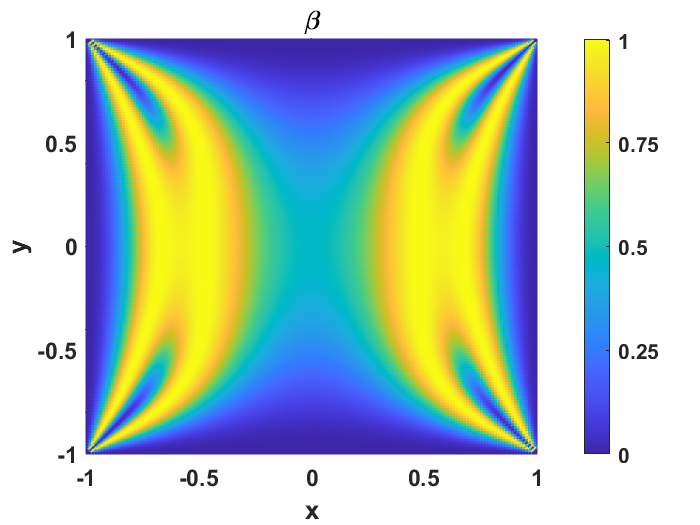}
    \end{minipage}
    \begin{minipage}{0.3\textwidth}
    \centering
        \includegraphics[width=1.0\textwidth]{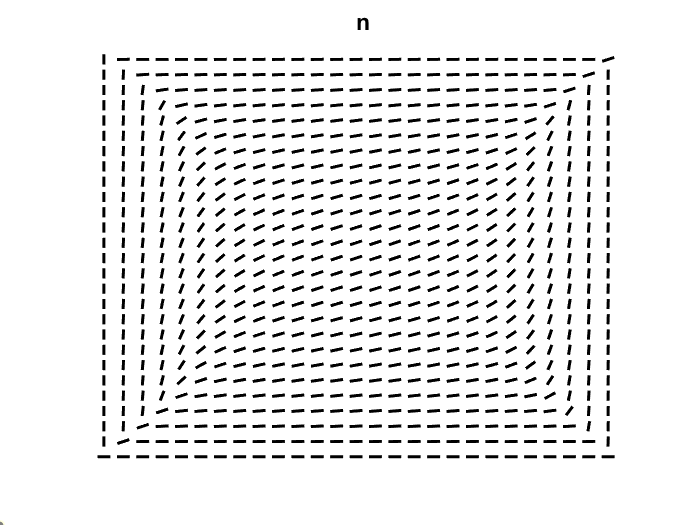}
    \end{minipage}
    \caption{First row: biaxiality parameter and director of an approximate WORS solution to \eqref{eq:full_stochastic} (under additive noise) for $\tilde{L}=0.05$, $\sigma=1$, $\alpha=3$ and $T=2$; $q_1(0,0)=0.0215$ for this solution. Second row: biaxiality parameter and director of an approximate BD$_x$ solution to \eqref{eq:full_stochastic} (under additive noise) for $\tilde{L}=0.05$, $\sigma=1$, $\alpha=3$ and $T=2$.} 
    \label{fig:lambda_0.05_alpha_3_WORS_full}
\end{figure}

\begin{figure}[ht]
    \centering
    \begin{minipage}{0.5\textwidth}
        \centering
        \includegraphics[width=1.0\textwidth]{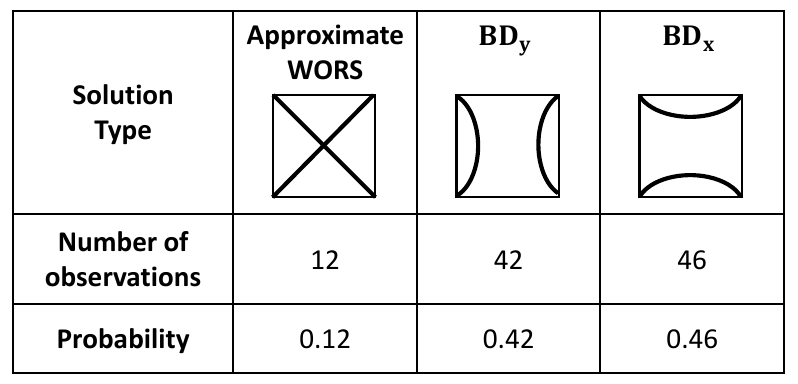}
    \end{minipage}
    \caption{Frequencies of observed solutions of \eqref{eq:full_stochastic} (under additive noise) in 100 simulations (with random initial conditions) for $\tilde{L}=0.05$, $\sigma=1$, $\alpha=3$ and $T=2$.} 
    \label{fig:frequencies_lam_0.05}
\end{figure}

For the remainder of this subsection, we study \eqref{eq:full_stochastic}. We first set $\alpha=3$, $L=0.05$ and use a WORS initial condition 
to check what profiles exist for the full system of equations. We still recover approximate WORS solutions. However, in contrast to \Cref{fig:lambda_0.05_alpha_3_WORS}, the $\pi/2$ change in the director orientation across the square diagonals is replaced by smooth rotation 
(see \Cref{fig:lambda_0.05_alpha_3_WORS_full} first row). This is a consequence of $q_2\neq0$ everywhere in $\Omega$ because of the inclusion of noise in \eqref{eq:q_2_stochastic}. Hence, to classify a solution as approximate WORS in this case, we additionally require that the average value of $|q_2|$ throughout $\Omega$ to be less than $0.05$. We also recover approximate BD solutions (see \Cref{fig:lambda_0.05_alpha_3_WORS_full} second row, again we have smooth rotation of the director between different regions of the square) 
for these parameter values and BD initial condition. We label a solutions as approximate BD if the average value of $|q_2|<0.2$ throughout $\Omega$, and the director and biaxiality plot qualitatively resemble a BD profile (see \Cref{fig:summary} second row for reference).
We complete 100 simulations with $\tilde{L}=0.05$, $\alpha=3$ (weak noise) and a random initial condition (the entries of $q_1^0$ and $q_2^0$ are generated from a uniform distribution on $[-1,1]$), and record the observed solutions in \Cref{fig:frequencies_lam_0.05} (the solutions are classified from each simulation by looking at the director and biaxiality plots).
The probability of observing an approximate WORS is 0.12, while the probability of observing an approximate BD solution is 0.88. Therefore, we deduce that BD solutions are the most relevant profiles for small square sizes under the inclusion of noise, and hence, most likely to be observed in experiments on small nano-scale domains.  

\begin{figure}[ht]
    \centering
    \begin{minipage}{0.3\textwidth}
        \centering
        \includegraphics[width=1.0\textwidth]{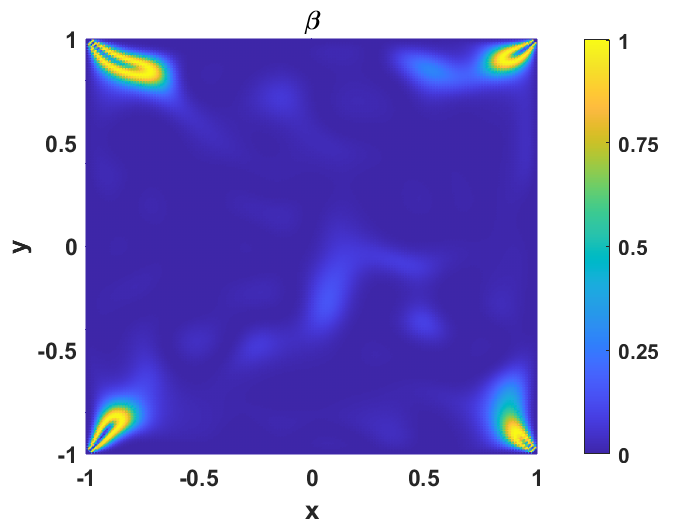}
    \end{minipage}
    \begin{minipage}{0.3\textwidth}
    \centering
        \includegraphics[width=1.0\textwidth]{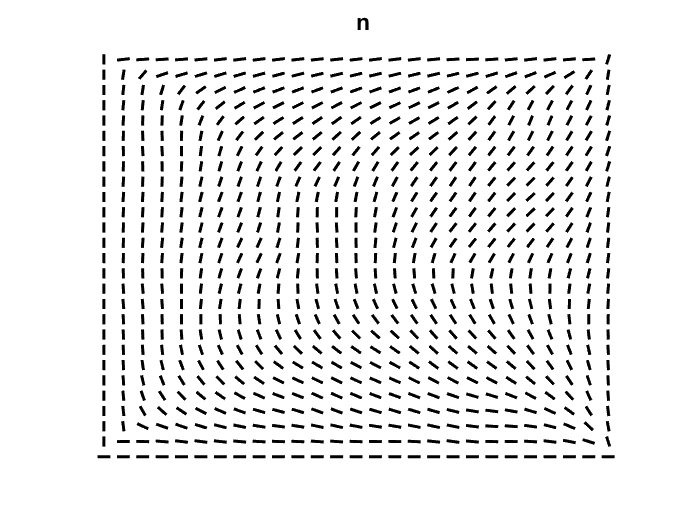}
    \end{minipage}
    \centering\\
    \begin{minipage}{0.3\textwidth}
        \centering
        \includegraphics[width=1.0\textwidth]{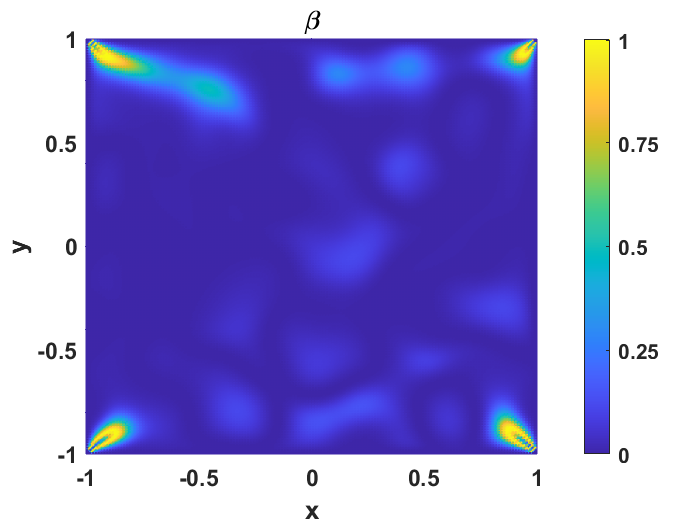}
    \end{minipage}
    \begin{minipage}{0.3\textwidth}
    \centering
        \includegraphics[width=1.0\textwidth]{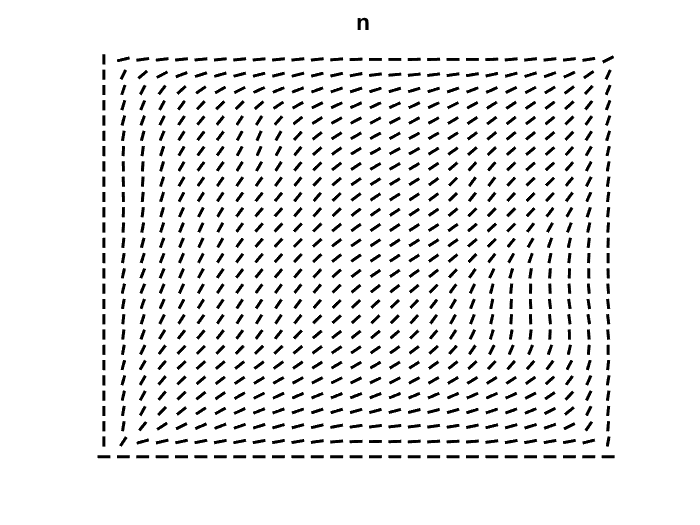}
    \end{minipage}
    \caption{First row: biaxiality parameter and director of a rotated solution to \eqref{eq:full_stochastic} (under additive noise) for $\tilde{L}=200$, $\sigma=1$, $\alpha=0.01$ and $T=2$. Second row: biaxiality parameter and director of a diagonal solution to \eqref{eq:full_stochastic} (under additive noise) for $\tilde{L}=200$, $\sigma=1$, $\alpha=0.01$ and $T=2$.} 
    \label{fig:lambda_200_alpha_0.01_rotated}
\end{figure}

\begin{figure}[ht]
    \centering
    \begin{minipage}{0.8\textwidth}
        \centering
        \includegraphics[width=1.0\textwidth]{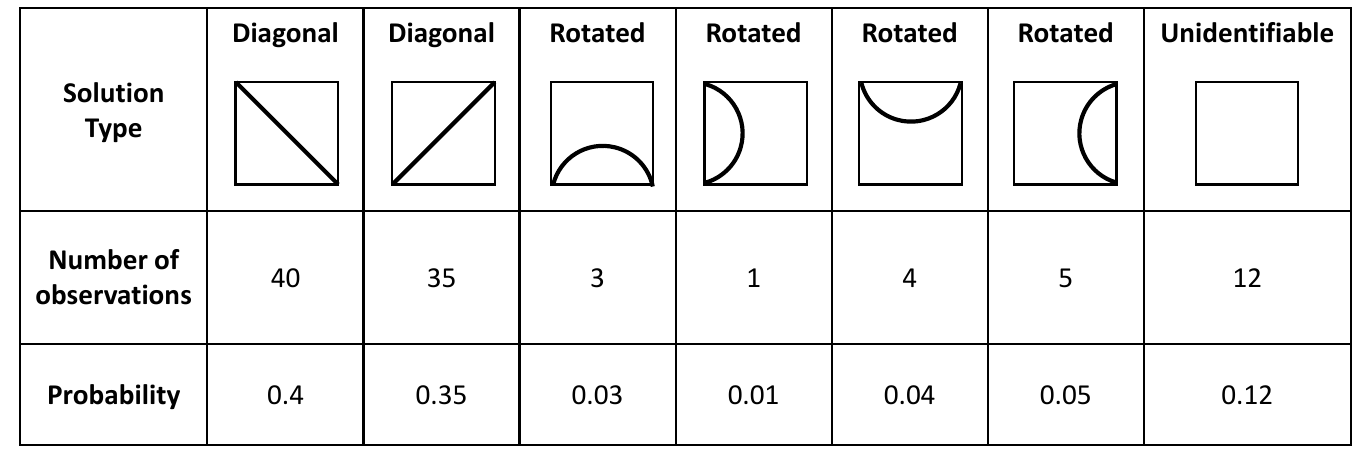}
    \end{minipage}
    \caption{Frequencies of observed solutions of \eqref{eq:full_stochastic} (under additive noise) in 100 simulations (with random initial conditions) for $\tilde{L}=200$, $\sigma=1$, $\alpha=0.01$ and $T=2$.} 
    \label{fig:frequencies_lam_200}
\end{figure}

Increasing $\tilde{L}$ or the square edge length, we are unable to find approximate WORS and approximate BD solutions for $\tilde{L}\geq 6.5$ and $\tilde{L}\geq 27.5$ respectively, in the context of the full stochastic system \eqref{eq:full_stochastic}, with weak additive noise $\alpha=3$ (that is, in 20 simulations we did not observe these solutions for these parameter values using a WORS and BD initial condition respectively). 
Using a diagonal initial condition, 
 diagonal solutions are found for $\tilde{L}\geq 7.5$ when $\alpha=3$ (in the sense that director and biaxiality plots are qualitatively similar to a diagonal solution from the deterministic setting); this deviates slightly from the deterministic approach where they are observed for $\tilde{L}\geq 6.4$. Using a rotated initial condition, we find rotated solutions for $\tilde{L}\geq 34$ with $\alpha=3$ (in the sense that director and biaxiality plots qualitatively resemble a rotated solution from the deterministic setting); this again deviates from the deterministic case for which rotated solutions are observed for $\tilde{L}\geq 28$. For large $\tilde{L}=200$, the observed diagonal and rotated solutions are essentially unperturbed by additive noise for $\alpha>0.1$ (relatively strong noise). In the case of strong noise with $\alpha=0.01$, the observed profiles (see \Cref{fig:lambda_200_alpha_0.01_rotated}) remain clearly identifiable with their deterministic counterparts.  
Completing 100 simulations with $\tilde{L}=200$, $\alpha=0.01$ and a random initial condition, diagonal solutions are observed far more frequently than rotated solutions, with a probability of 0.75 (see \Cref{fig:frequencies_lam_200}) and as such, we deduce that diagonal solutions are the most likely to be observed in experiments. This is in agreement with the energy calculations in the deterministic Oseen-Frank framework in \cite{lewis_experiments}, that demonstrate that the diagonal solutions have lower energy than rotated solutions and lower energy solutions are more likely to be observed in experiments.

\subsubsection{Effects of multiplicative noise}
This subsection focuses on the addition of multiplicative noise to the stochastic PDEs \eqref{eq:q_1_stochastic}, \eqref{eq:q_2_stochastic} i.e., $dW$ is multiplied by a function which depends on the unknown being solved for \cite{into_book}. Specifically, we let 
\begin{equation}\label{eq:multiplicative_noise}
    G_1(q_1,q_2)=q_1 \textrm{ and } G_2(q_1,q_2)=q_2,
\end{equation}
in \eqref{eq:q_1_stochastic} and \eqref{eq:q_2_stochastic}, respectively. This choice of the multiplicative noise does not affect points where $q_1=0$ or $q_2=0$, so it preserves symmetric structures or defect lines with $q_1 = q_2 = 0$. We have chosen the multiplicative factors of $q_1,\;q_2$, instead of some other functions $G_1(q_1,q_2),\;G_2(q_1,q_2)$, in an attempt to observe/stabilise the WORS in a stochastic setting. 
Using a WORS initial condition, 
this multiplicative noise helps preserve $q_1=0$ on the square diagonals and $q_2=0$ everywhere, and hence the symmetry of the WORS. Physically, the preservation of points with $q_1=0$ and $q_2=0$, would correspond to some kind of strong surface treatment which fixes molecules to lie in certain directions in the square interior. 

We only consider \eqref{eq:full_flow} with the addition of multiplicative noise, and not \eqref{eq:q_flow}. Since noise is not introduced at points where $q_2=0$ in the multiplicative case, with appropriate initial conditions (i.e., $q^0_2=0$) one can find solutions of \eqref{eq:q_1_stochastic}, \eqref{eq:q_2_stochastic} with $q_2=0$ in $\Omega$, making a study of \eqref{eq:stochastic_q} with multiplicative noise redundant. In this section,  $\sigma=1$ is fixed and $\alpha$ is varied to control the strength of the noise.

\begin{figure}[ht]
    \centering
    \begin{minipage}{0.32\textwidth}
        \centering
        \includegraphics[width=1.0\textwidth]{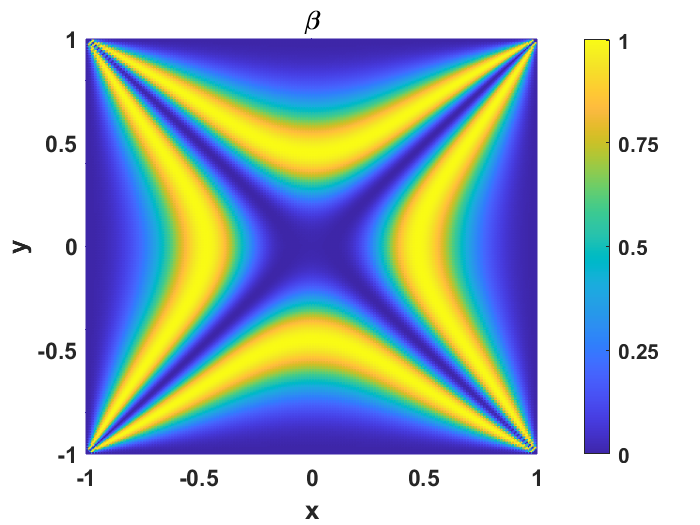}\\
    \end{minipage}
    \begin{minipage}{0.32\textwidth}
    \centering
        \includegraphics[width=1.0\textwidth]{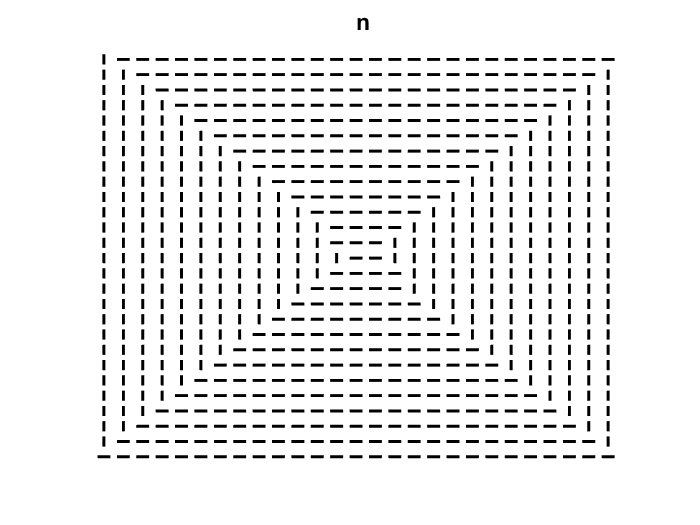}\\
    \end{minipage}
    \caption{Biaxiality parameter and director of an approximate WORS solution to \eqref{eq:full_stochastic} (under multiplicative noise) for $\tilde{L}=10$, $\sigma=1$, $\alpha=1$ and $T=2$; $q_1(0,0)=0.015$ for this solution.} 
    \label{fig:lambda_10_alpha_1_approx_WORS_multiplicative}
\end{figure}

For the system \eqref{eq:q_1_stochastic}, \eqref{eq:q_2_stochastic} under multiplicative noise, with a WORS initial condition, 
we are able to find WORS solutions (as characterised in \Cref{sec:deterministic_numerics}), as well as approximate WORS solutions for larger values of $\tilde{L}$ and $\alpha$ than we could under additive noise. In \Cref{fig:lambda_10_alpha_1_approx_WORS_multiplicative} for instance, we present an approximate WORS solution of \eqref{eq:full_stochastic} for $\tilde{L}=10$ and $\alpha=1$. In \Cref{table:q_values}, we record values of $q_1(0,0)$ for different $\Tilde{L}$ and $\alpha$ (we record the smallest value of $q_1(0,0)$ seen in 10 simulations for a given choice of $\Tilde{L}$ and $\alpha$), hence summarising when we observe WORS and approximate WORS solutions of \eqref{eq:full_stochastic}. Recall a WORS solution exists for all $\tilde{L}\geq 0$ in the deterministic case. For multiplicative noise with $\alpha=3$, we do not find WORS solutions for $\tilde{L}\geq 14$ and approximate WORS solutions for $\Tilde{L}\geq 28$, while for additive noise with $\alpha=3$, we could not find WORS solutions for any value of $\tilde{L}$ and approximate WORS solutions for $\Tilde{L}\geq 6.5$ (in the context of either \eqref{eq:stochastic_q} or system \eqref{eq:full_stochastic}). Clearly the introduction of multiplicative noise governed by \eqref{eq:multiplicative_noise} enhances the stability of WORS and approximate WORS solutions. Completing 100 simulations with $\tilde{L}=0.05$, $\alpha=3$ and a random initial condition, 94 WORS solutions and 6 approximate WORS solutions are observed. This multiplicative noise also supports the observation of BD solutions, which are now observed for all $\tilde{L}>0$ using a BD initial condition and are visually qualitatively identical to the BD solutions in the deterministic case. As such, they are not presented. This is significantly different to the additive case, for which we could not find BD solutions for $\tilde{L}\geq 27.5$, for the full system \eqref{eq:full_stochastic}.

\begin{table}
\begin{center}
 \begin{tabular}{|c | c| c | c |}
 \hline
   & $\tilde{L}=0.05$ & $\tilde{L}=1$ & $\Tilde{L}=10$\\  
 \hline
 $\alpha=0.01$ & 0.0079  & -0.0110 & 0.1083 \\
 \hline
 $\alpha=0.1$ & $9.4971\times 10^{-4}$  & -0.0029  & -0.0778 \\
 \hline
 $\alpha=1$ & $8.0814\times 10^{-6}$ & $3.7377\times 10^{-5}$ & 0.0011 \\
 \hline
 $\alpha=3$ & $-4.5992\times 10^{-10}$ & $-2.4608\times 10^{-9}$ & $-1.4024\times 10^{-7}$\\
 \hline
\end{tabular}
\end{center}
\caption{\label{table:q_values} Value of $q_1(0,0)$ obtained for solutions to \eqref{eq:full_stochastic} (under multiplicative noise), for different values of $\alpha$ and $\Tilde{L}$. If $|q_1(0,0)|\leq 10^{-6}$ and the average value of $|q_1|\leq 10^{-6}$ on the square diagonals (this is true in the table above but the values are not presented), it is a WORS solution, if $10^{-6}<|q_1(0,0)|<0.05$ and the average value of $10^{-6}<|q_1|<0.05$ on the square diagonals (this is true in the table above but the values are not presented), it is approximate WORS, and BD otherwise. For WORS solutions, the average value of $|q_2|$ throughout $\Omega$ is less than $10^{-6}$ and less than 0.05 for approximate WORS solutions.}
\end{table}

Using a diagonal initial condition, diagonal solutions are found for $\tilde{L}\geq 5.4$, when $\alpha=3$.
For comparison, in the deterministic case and in the stochastic case with additive noise (with $\alpha=3$), we observe diagonal solutions for $\tilde{L}\geq 6.4$ and $\tilde{L}\geq 7.5$ respectively. With a rotated initial condition, 
we are also able to find rotated solutions for $\tilde{L}\geq 25$. Again, for comparison, in the deterministic case and in the stochastic case with additive noise (with $\alpha=3$), we observe rotated solutions for $\tilde{L}\geq 28$ and $\tilde{L}\geq 34$ respectively. Hence, the multiplicative noise \eqref{eq:multiplicative_noise} preserves solutions with $q_2 \equiv 0$ for larger values of $\tilde{L}$, or larger square domains, than in the additive case, without hindering the observation of solutions with $q_2\neq0$ such as the diagonal and rotated solutions. 
In particular, in the case of multiplicative noise, we find that solutions with $q_2=0$ everywhere in $\Omega$ (BD solutions and WORS) can co-exist with solutions that have $q_2\neq 0$ everywhere in $\Omega$ (diagonal and rotated solutions), for the same domain size, and hence multiplicative noise enhances multistability and also the probability of observing solutions with interior biaxiality such as the WORS or the BD solutions.

\begin{figure}[ht]
    \centering
    \begin{minipage}{0.8\textwidth}
        \centering
        \includegraphics[width=1.0\textwidth]{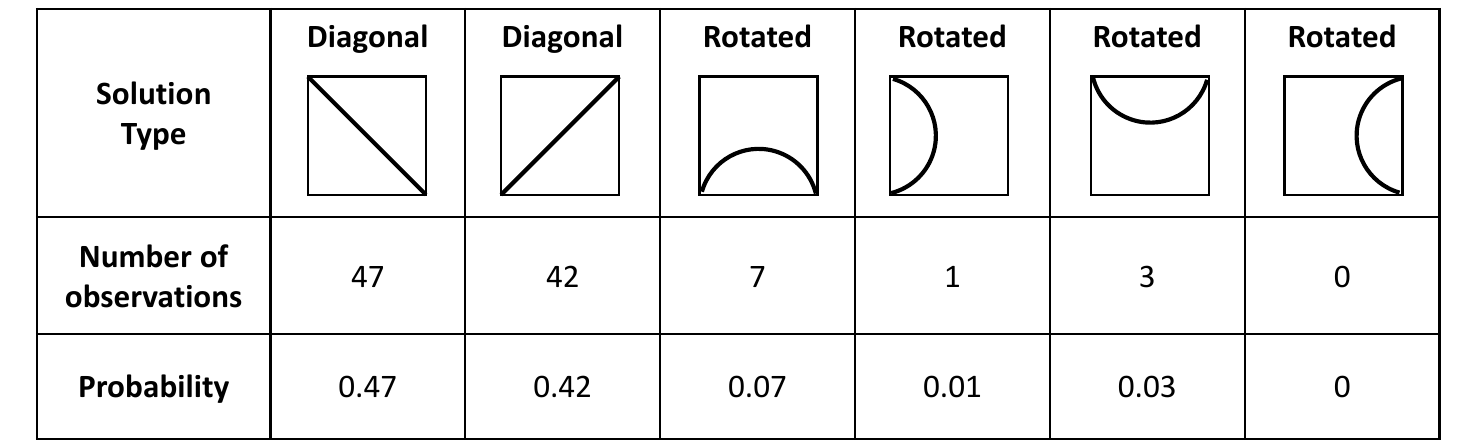}
    \end{minipage}
    \caption{Frequencies of observed solutions of \eqref{eq:full_stochastic} (under multiplicative noise) in 100 simulations (with random initial conditions) for $\tilde{L}=200$, $\sigma=1$, $\alpha=0.01$ and $T=2$.} 
    \label{fig:frequencies_lam_200_multiplicative}
\end{figure}

We complete 100 simulations with $\tilde{L}=200$, $\alpha=0.01$ (strong noise) and a random initial condition and record the observed solutions in \Cref{fig:frequencies_lam_200_multiplicative}. The random initial conditions excludes BD-type solutions for this large square domain.
Diagonals solutions are again observed far more frequently than rotated solutions, with a probability of 0.89, further supporting that they are the most relevant solution for large square sizes. Furthermore, all solutions in these 100 simulations are identifiable with a deterministic counterpart (unlike \Cref{fig:frequencies_lam_200}), implying that the impact of noise is less pronounced in the multiplicative case.


\subsubsection{Potential applications to switching processes}\label{sec:switching}

The inclusion of noise may be used to model the switching processes in a multistable liquid crystal device. The key principle in such devices, is that an external input (e.g. applied electric or magnetic fields) is needed only to switch between distinct equilibria or energy minimisers, but not necessarily to maintain individual states, thus offering the prospect of power-efficient and high resolution optical devices. 
Mathematically, the switching process is described by a non-equilibrium time-dependent process, that drives the system out of a given critical point of \eqref{eq:energy} and pushes the system into a different critical point of \eqref{eq:energy}, by overcoming their energy barrier (i.e., the difference between the value of \eqref{eq:energy} for each critical point). The switching is typically mediated by external inputs, which could be an external electric/magnetic field, thermal effects or fluctuations, and the input  essentially provides a kick to the system, and once the input is removed, the system settles into a different equilibrium configuration. To model large fluctuations that could induce a switching process, we take a solution of the deterministic system \eqref{eq:EL-equations} as our initial condition and introduce additive noise (i.e. consider\eqref{eq:full_stochastic} with $G_1=G_2=1$) of a given strength $\alpha$, at our first time step. The value of $\alpha$ or the noise strength needed to facilitate switching, also depends on the initial condition. 
We keep the value of $\alpha$ fixed until $T=0.1$, at which point we increase $\alpha=10$, to weaken the noise and remove large fluctuations that could induce a switching process. We then further run our simulation until $T=0.5$, at which point our stochastic solution appears to converge to a different deterministic solution, simulating a switching process between the initial condition and the final solution.

\begin{figure}[ht]
    \centering
    \begin{minipage}{1.0\textwidth}
        \centering
        \includegraphics[width=1.0\textwidth]{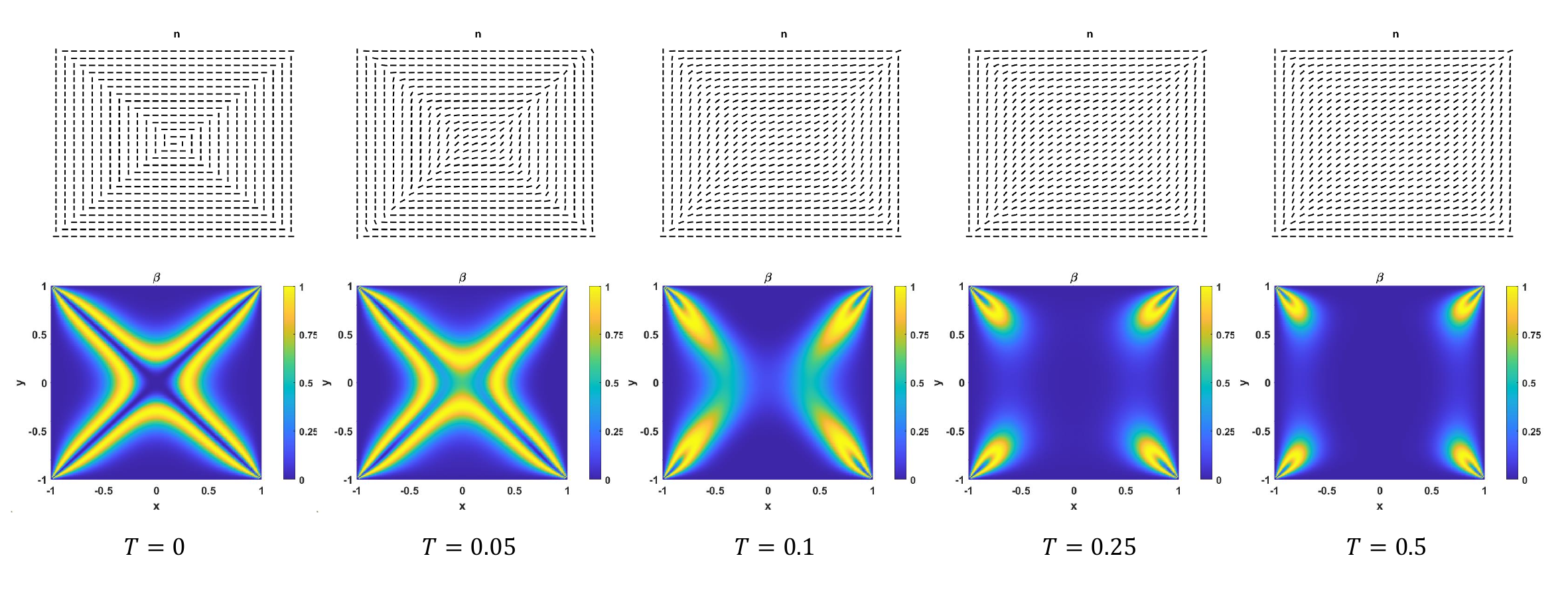}
    \end{minipage}
    \caption{Switching process from a (deterministic) WORS solution to a (stochastic) diagonal solution for $\tilde{L}=30$.} 
    \label{fig:switch_WORS_L_30}
\end{figure}

\begin{figure}[ht]
    \centering
    \begin{minipage}{1.0\textwidth}
        \centering
        \includegraphics[width=1.0\textwidth]{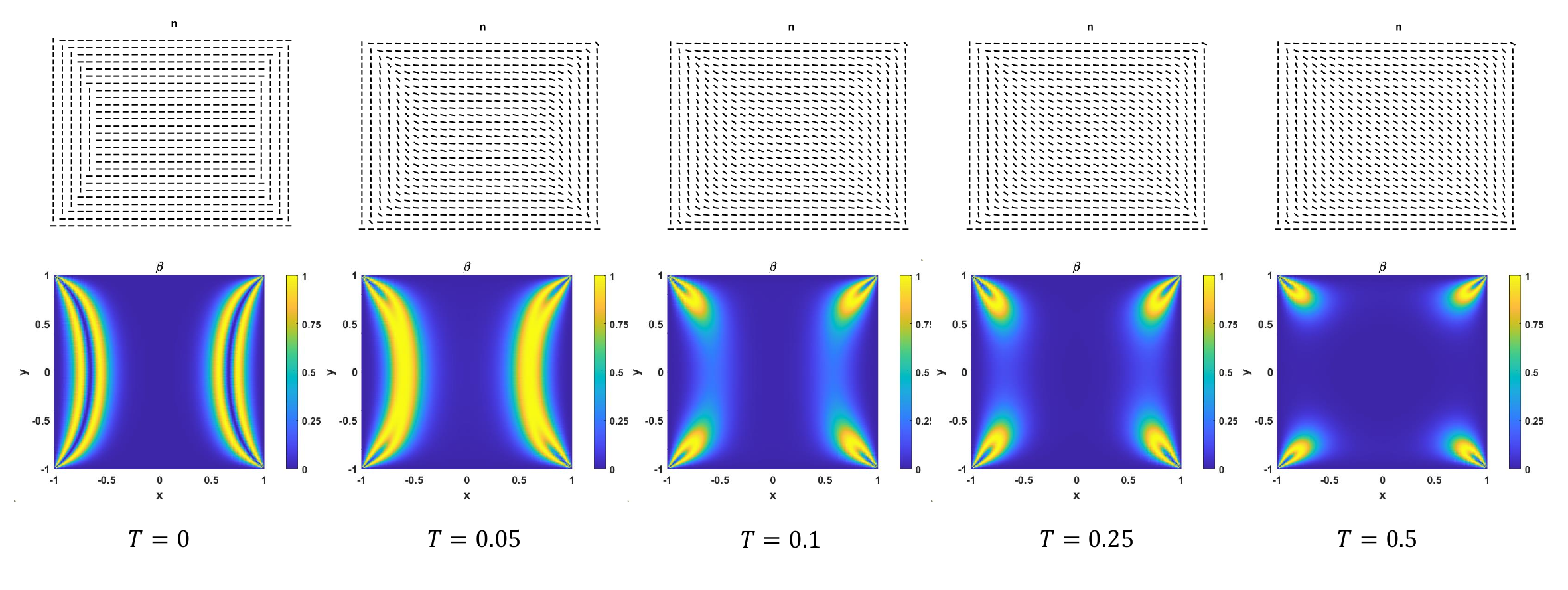}
    \end{minipage}
    \caption{Switching process from a (deterministic) BD solution to a (stochastic) diagonal solution for $\tilde{L}=30$.} 
    \label{fig:switch_BD_L_30}
\end{figure}

\begin{figure}[ht]
    \centering
    \begin{minipage}{1.0\textwidth}
        \centering
        \includegraphics[width=1.0\textwidth]{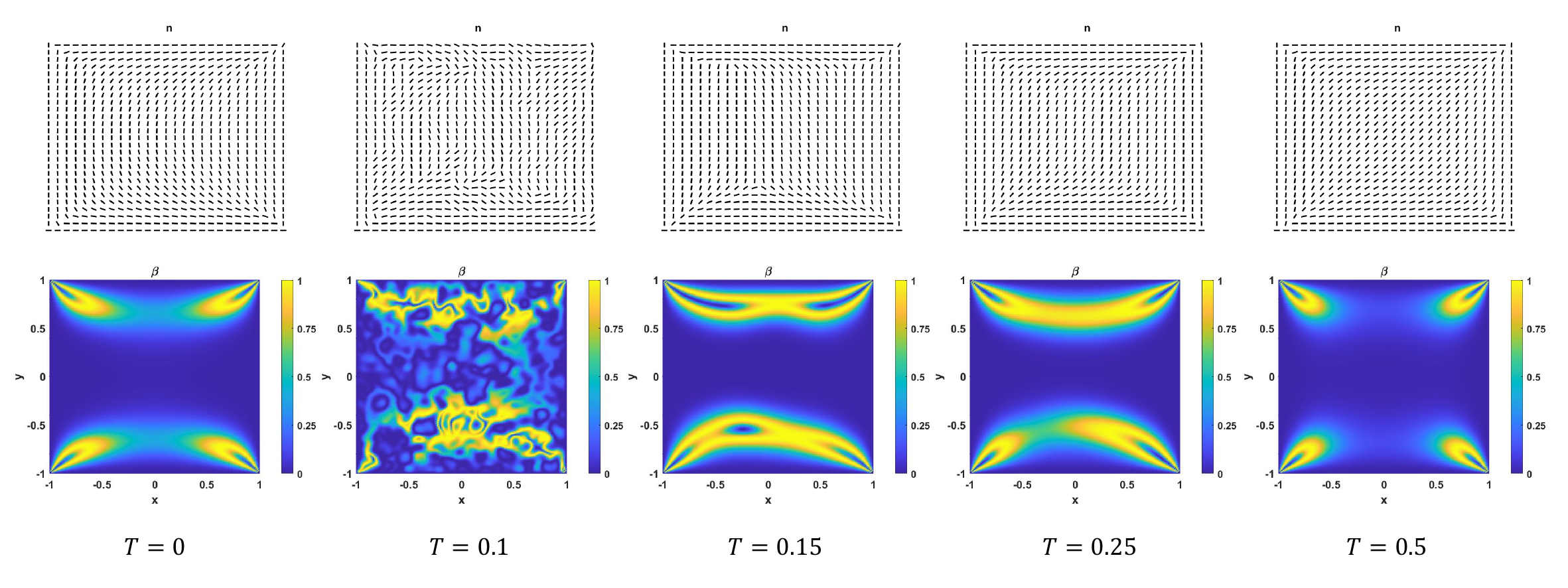}
    \end{minipage}
    \caption{Switching process from a (deterministic) rotated solution to a (stochastic) diagonal solution for $\tilde{L}=30$.} 
    \label{fig:switch_rotated_L_30}
\end{figure}

Some of our results that follow from this procedure are summarised below. These results are not comprehensive, but illustrate some generic concepts.
\begin{itemize}
    \item For $\tilde{L}=5$ and with the WORS as an initial condition, the system evolves to an approximate BD solution with $\alpha=3$ (weak noise).

    \item For $\tilde{L}=30$ and starting from a WORS, BD or rotated solution, the system evolves to a diagonal solution. 
    To make the system switch, we require $\alpha=3$ (weak noise) for the WORS initial condition, $\alpha=1$ (intermediate strength noise) for BD initial condition and $\alpha=0.001$ (very strong noise) for rotated initial conditions. Some snapshots of the switching processes from the WORS to diagonal, BD to diagonal and rotated to diagonal state are shown in \Cref{fig:switch_WORS_L_30}, \Cref{fig:switch_BD_L_30} and \Cref{fig:switch_rotated_L_30} respectively.

    \item For $\tilde{L}=200$ and with a WORS or BD initial condition, the system evolves to a diagonal solution. Weak noise e.g. $\alpha=3$ is sufficient to facilitate the WORS to diagonal switching, whereas we need $\alpha=1$ to induce the BD to diagonal switching.

    \item For $\tilde{L}=200$ and with a rotated solution, and very strong noise ($\alpha=0.001$), we find that in 4 out of 10 simulations the system moves to a diagonal solution, while in the remaining simulations, the system stays in the rotated solution.

    \item Our numerical simulations suggest that it is not possible to switch from a diagonal initial condition for square domains of any size, e.g. with $\tilde{L}=10,30,200$, and with very strong noise ($\alpha=0.001$), we cannot switch from a diagonal solution. This would be consistent with previous studies which suggest that the diagonal solution is the global energy minimiser for large square domains, with tangent boundary conditions. 

\end{itemize}

\section{The radial hedgehog}\label{sec:droplet}

To conclude this work, we perform some preliminary studies of the effects of additive noise on the radial hedgehog solution, on spherical droplets of nematic liquid crystal \cite{gartland_hedgehog}, with homeotropic boundary conditions i.e. a Dirichlet boundary condition of the form
\[
\Qvec_b = s_+ \left(\frac{\mathbf{r}\otimes \mathbf{r}}{r^2} - \frac{\mathbf{I}}{3} \right)
\]
where $s_+$ is defined in Section~\ref{sec:nematic_square}, $\mathbf{r}$ is the position vector and $\mathbf{I}$ is the $3 \times 3$ identity matrix.. The computational domain is a spherical droplet of radius, $\tilde{R}$:
\begin{equation}
    \tilde{\Omega}:=B(0,\tilde{R})=\{\mathbf{r}\in\mathbb{R}^3:|\mathbf{r}|\leq \tilde{R}\}.
\end{equation}
Consider the free energy \eqref{eq:dimensional_energy}, which can be non-dimensionalised to yield the dimensionless free energy:
\begin{equation}\label{eq:radial_energy}
    F(\Qvec)=\int_{B(0,R)} \frac{1}{2}|\nabla\Qvec|^2+\frac{t}{2}\textrm{tr}(\Qvec^2) -\sqrt{6}\textrm{tr}(\Qvec^3) +\frac{1}{2}(\textrm{tr}\Qvec^2)^2~\mathrm{d}\Omega,
\end{equation}
where $t_{temp}=27AC/B^2<0$ (is a measure of the system temperature) and $R=\Tilde{R}/\xi$, $\xi=\sqrt{27C/LB^2}$ being the biaxial correlation length \cite{virga}.
The radial hedgehog is a uniaxial radially symmetric solution of the corresponding Euler-Lagrange equations of \eqref{eq:radial_energy}, subject to the specified homeotropic Dirichlet boundary conditions. More formally, the radial hedgehog solution can be written in the form
\begin{equation}
    \Qvec(r)=\sqrt{\frac{3}{2}}h(r)\left(\frac{\mathbf{r}}{|\mathbf{r}|}\otimes\frac{\mathbf{r}}{|\mathbf{r}|}-\frac{1}{3}\mathbf{I}\right),\quad r\in[0,R],
\end{equation}
where $\mathbf{r}$ is the radial vector and $h$ is a solution of the ordinary differential equation
\begin{equation}\label{eq:h_eqtn}
    \frac{d^2 h}{dr^2} + \frac{2}{r}\frac{dh}{dr}-\frac{6h}{r^2}=t_{temp}h -3h^2 + 2h^3, \quad r\in(0,R)
\end{equation}
subject to
\begin{equation}
    h(0)=0 \textrm{ and } h(R)=h_+:=\frac{3+\sqrt{9-8t_{temp}}}{4}.
\end{equation}
Here, $h_+$ is the re-scaled definition of $s_+$ in \eqref{eq:s_+}. The radial hedgehog solution is fully defined by the solution of the ordinary differential equation \eqref{eq:h_eqtn}, subject to the fixed boundary conditions. 

For implementing noise in \eqref{eq:h_eqtn}, one option is to include one-dimensional additive noise in \eqref{eq:h_eqtn} and solve the equation independently in every radial direction, yielding the full profile of the droplet. However, for simplicity, we solve \eqref{eq:h_eqtn} on a circular cross section of the sphere (i.e., a disk), enabling us to implement the two-dimensional $Q$-Wiener process outlined in \cref{sec:stochastic_setup}. That is, we specify $N$ independent copies of equation \eqref{eq:h_eqtn} corresponding to $N$ radial directions on the disk, making the problem effectively two-dimensional, and then add the two-dimensional noise seen previously. Points on the disk can be specified using polar coordinates $(r,\theta)$, where $r\in[0,R]$ and $\theta\in[0,2\pi]$. Hence, to generate noise, we consider this as a problem on the rectangle $[0,R]\times[0,2\pi]$, making the implementation the same 
 as in \Cref{sec:stochastic_setup} after adjusting for the different interval widths in Algorithm 10.5 \cite{into_book}.   
We therefore solve the following equation 
\begin{equation}\label{eq:h_eqtn_noise}
    dh=\left[\frac{d^2 h}{dr^2} + \frac{2}{r}\frac{dh}{dr}-\frac{6h}{r^2}-t_{temp}h +3h^2 - 2h^3\right]dt +\sigma dW, \;t\in[0,T],
\end{equation}
on $[0,R]\times [0,2\pi]$, which is divided into a $100\times 100$ grid of points and again consider solutions at $T=2$, with $\Delta t=2\times 10^{-5}$ as our time step. We then present the average of the $N=100$ $h$ profiles obtained on the disk, to assess the impact of noise on the radial hedgehog profile. The initial condition is fixed to be $h=r$, for $r\in[0,R]$ throughout this section, but $t_{temp}$, $R$ and $\alpha$ are varied to study the effects of the model parameters and noise on the radial hedgehog solution. In the definition of $t_{temp}$, we fix $B=6400$Nm$^{-2}$ and $C=3500$Nm$^{-2}$ \cite{majumdar-2010-article}, so that varying $t_{temp}$ is equivalent to varying $A$ and the system temperature. Recall that the incorporation of noise is intended to capture material imperfections and experimental uncertainty. 


\begin{figure}[ht]
\centering
    \begin{minipage}{0.45\textwidth}
        \centering
        \includegraphics[width=1.0\textwidth]{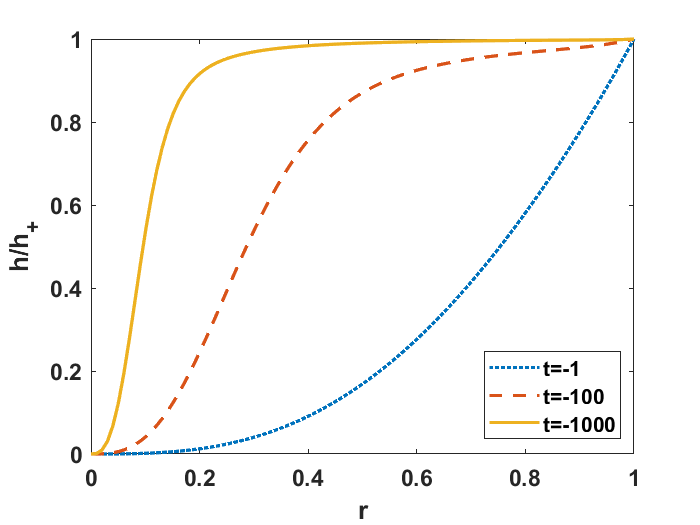}\\
    \ \textrm{(a)}
    \end{minipage}
    \centering
    \begin{minipage}{0.45\textwidth}
        \centering
        \includegraphics[width=1.0\textwidth]{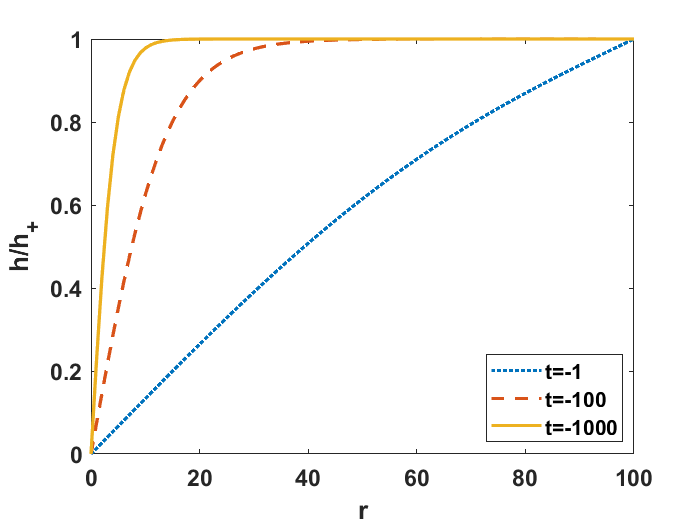}
        \\
    \ \textrm{(b)}
    \end{minipage}
    \caption{Deterministic solution $h$, of \eqref{eq:h_eqtn} on $[0,R]$, with $t_{temp}$ (marked as $t$) as indicated, $R=1$ (a) and $R=100$ (b). $t_{temp}=-1,-100,-1000$ correspond to $A=-433$Nm$^{-2}$, $-4.33\times 10^{4}$Nm$^{-2}$, $-4.33\times 10^{5}$Nm$^{-2}$, respectively.} 
    \label{fig:deterministic_h_sol}
\end{figure}

In \Cref{fig:deterministic_h_sol}, we numerically compute solutions of the deterministic equation \eqref{eq:h_eqtn} (with no noise) for $R=1,100$ and multiple values of $t_{temp}$. In \Cref{fig:deterministic_h_sol} (a), as $t_{temp}$ decreases (for $R=1$), deep into the nematic phase, the degree of interior nematic ordering increases since $h/h_+\to 1$ \cite{majumdar_2012_radialhedgehog,lamy_hedgehog}. Similarly, a larger value of $R$ e.g. $R=100$ has the same effect of increasing the interior ordering, for a fixed value of $t_{temp}$, compared to the the $R=1$ case. 

\begin{figure}[ht]
\centering
    \begin{minipage}{0.45\textwidth}
        \centering
        \includegraphics[width=1.0\textwidth]{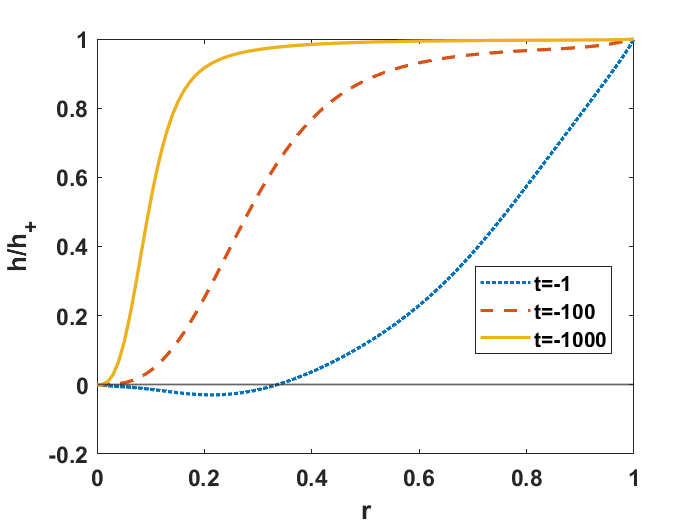}\\
    \ \textrm{(a)}
    \end{minipage}
    \centering
    \begin{minipage}{0.45\textwidth}
        \centering
        \includegraphics[width=1.0\textwidth]{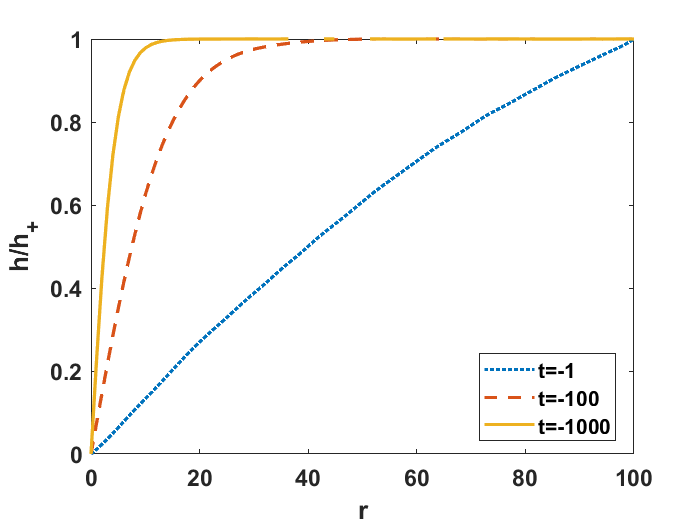}
        \\
    \ \textrm{(b)}
    \end{minipage}
    \caption{Average of 100 stochastic solutions $h$, of \eqref{eq:h_eqtn_noise} on a disk, with $T=2$, $\sigma=1$, $t_{temp}$ (marked as $t$) as indicated and (a) $R=1$, $\alpha=0.5$, (b) $R=100$, $\alpha=0.01$. }
    \label{fig:stochastic_h_sol}
\end{figure}

For comparison, in \Cref{fig:stochastic_h_sol}, we plot the average of $100$ solutions of the stochastic equation \eqref{eq:h_eqtn_noise}, for $R=1,100$, the same values of $t_{temp}$ as in \Cref{fig:deterministic_h_sol}, and with $\sigma=1$, $\alpha=0.5,0.01$ and $T=2$. We consider $\alpha=0.5$ in \Cref{fig:stochastic_h_sol} (a), to be an intermediate noise strength.
Interestingly, in \Cref{fig:stochastic_h_sol} (a), with $R=1$ and $t_{temp}=-1$, we find $h/h_+<0$ on an interval $(0,r_n)$ where $r_n\approx 0.35$. This is important, because it demonstrates that the radial symmetry of the radial hedgehog solution could be violated (which requires non-negative $h$) for sufficiently small droplets with small negative values of $t$, or equivalently high temperatures ($t_{temp}=-1$ corresponds to $A=-433$Nm$^{-2}$). For large negative values of $t_{temp}$ (i.e, $t_{temp}<-100$) deep in the nematic phase, the inclusion of noise has no discernible impact on the $h/h_+$ profiles, which look visually similar to the plots in \Cref{fig:deterministic_h_sol} (a). 
This is in agreement with experiments, where radially symmetric optical patterns are seen at sufficiently low temperatures \cite{experiments}. In \Cref{fig:stochastic_h_sol} (b), where $R=100$ and $\alpha=0.01$ (i.e., strong noise), noise has no noticeable impact on the scalar order parameter (on average). This suggests, as in the case of square domains, that noise has less effect on solution profiles for large domains.

Performing numerical experiments for different values of $t_{temp}$, $\alpha$ and $R$, we find:
\begin{itemize}
    \item for $R=1$ and $t_{temp}=-1$, $\alpha=0.75$ is the weakest noise required to observe $h<0$ (on average in 100 simulations);
    
    \item for $R=1$ and $\alpha=0.5$, $t_{temp}=-1.2$ ($A=520$Nm$^{-2}$) is the lowest temperature for which $h<0$ can be observed (on average in 100 simulations), suggesting that the critical noise needed for symmetry breaking increases with decreasing temperature;
    
    \item and finally, for $t_{temp}=-1$ and $\alpha=0.5$, $R=1.1$ is the largest droplet size for which $h<0$ can be observed (on average in 100 simulations). This suggests that for a fixed $t_{temp}, \alpha$, there is a critical droplet size, $R_c(t_{temp}, \alpha)$ such that $h$ is non-negative for $R > R_c(t_{temp}, \alpha)$.
\end{itemize}
These numerical experiments suggest that symmetry-breaking may occur for the radial-hedgehog solution, for sufficiently small droplets and for sufficiently high temperatures, under the influence of additive noise. A negative value of $h$ implies that the molecules, on average, lie in the plane orthogonal to the radial unit-vector and this changes the defect profile near the droplet centre. In the deterministic case, the spherically symmetric radial hedgehog solution, with a unique, monotonic and non-negative $h$-profile, is globally stable for small droplets and for sufficiently high temperatures \cite{majumdar_2012_radialhedgehog,lamy_hedgehog}, in contrast to the stochastic predictions in this section. It is experimentally difficult to zoom into director profiles near the droplet centre, or defect profiles in general, and hence, it may be hard to test whether the deterministic predictions are indeed valid for small droplets and/or high temperatures, in real-life experimental settings which inevitably have some fluctuations and imperfections. 

\section{Conclusions}\label{sec:conclusions}

In this paper, we perform some numerical explorations of well-studied model problems in the Landau-de Gennes theory for nematic liquid crystals, with the effects of random noise. The random noise models material imperfections or uncertainties in the experimental set-up. The inclusion of random noise can be used as a method to test the robustness and physical relevance of solutions computed with deterministic approaches. We argue that solutions which survive (qualitatively) under the inclusion of noise are more likely to be observed in experiments than those which are not. For instance, with additive noise (including strong noise), diagonal and rotated solutions survive on large domains, suggesting they are the most physically relevant. This is consistent with experimental results which support the observation of diagonal and rotated solutions (see \cite{tsakonas_brown_davidson_mottram} and \cite{lewis_experiments} for instance). It also suggests that the deterministic approach adequately captures structural details in the parameter regime of large $\tilde{\lambda}$, when bulk effects dominate elastic effects. 

The picture is different for small domains. The WORS is the unique energy minimiser, and globally stable for small domains, in the deterministic Landau-de Gennes framework. The critical square edge length $\tilde{L}_c$ depends on the temperature, such that the WORS is globally stable for $\tilde{L} < \tilde{L}_c$. The WORS does not really survive with additive noise, although we observe approximate WORS for sufficiently small square domains. This suggests that the WORS maybe an artefact of the symmetries of the deterministic model, with a highly symmetric diagonal defect cross, and is perhaps why it has not been observed experimentally.
Hence, further effects must be included to enhance the stability of the WORS so it is consistently observed in a stochastic setting. In turn, this may inform the design of new experiments, to facilitate the observation of the WORS. 
The physical relevance of the multiplicative noise setup is questionable, as it requires preferred directions of molecular alignment to be enforced inside the square domain. 

Finally, the work in \Cref{sec:switching}, proposes a method for modelling switching processes via the inclusion of noise. We can clearly capture the switching mechanism from less stable to more stable solutions. However, switching between different energetically degenerate solutions e.g. from one diagonal solution to another, cannot be captured with this method. Future work on stochastic liquid crystal models can include:
\begin{enumerate}
    \item study of alternative numerical schemes such as Galerkin finite element methods for liquid crystal problems (i.e., systems of nonlinear partial differential equations);

    \item rigorous convergence analysis of numerical schemes for stochastic PDEs and their long time behaviour;
    
    \item generalisation of the approaches in this paper to study three-dimensional liquid crystal problems, for which the LdG $\Qvec$-tensor has five degrees of freedom, with emphasis on how noise affects defect structures and their stability. The stochastic study can also be used to test the robustness of deterministic predictions. 

  
\end{enumerate}

\section*{Acknowledgements}
The authors would like to thank Professor Neela Nataraj for her valued suggestions and feedback on the paper, especially with regards to the numeral implementation.  The authors also thank Professor Utpal Manna (IISER Trivandrum) for helpful references and discussions in the initial stages of the work.

\section*{Disclosure statement}
The authors report there are no competing interests to declare.

\section*{Funding}
AM is supported by the University of Strathclyde New Professors Fund, a Leverhulme
Research Project Grant RPG-2021-401, an OCIAM Visiting Fellowship at the University of Oxford and
a Daiwa Foundation Small Grant. The authors gratefully acknowledge funding from the Royal Society International Exchange Grant IES\textbackslash R2\textbackslash 202068. JD's PDRA is funded by the EPSRC Additional Funding for Mathematical Sciences scheme.


\section{References}

\bibliographystyle{tfnlm}
\bibliography{main}

\begin{thebibliography}{10}
\providecommand{\url}[1]{\normalfont{#1}}
\providecommand{\urlprefix}{Available from: }

\bibitem{deGennes}
de~Gennes~PG, Prost~J. {The Physics of Liquid Crystals}. 2nd ed. Oxford:
  Clarendon Press; 1993.

\bibitem{canevari_majumdar_spicer2017}
Canevari~G, Majumdar~A, Spicer~A. {Order reconstruction for nematics on squares
  and hexagons: a Landau-de Gennes study}. SIAM Journal on Applied Mathematics.
  2017;\hspace{0pt}77(1):267--293.

\bibitem{2D_landscape}
Robinson~M, Luo~C, Farrell~PE, et~al. {From molecular to continuum modelling of
  bistable liquid crystal devices}. Stochastic partial differential equations :
  analysis and computations. 2017;\hspace{0pt}44(14-15):2267--2284.

\bibitem{gartland_hedgehog}
Gartland~EC, Mkaddem~S. {Instability of radial hedgehog configurations in
  nematic liquid crystals under Landau–de Gennes free-energy models}. Phy Rev
  E. 1999;\hspace{0pt}59(1):563--567.

\bibitem{into_book}
Lord~GJ, Powell~CE, Shardlow~T. {An Introduction to Computational Stochastic
  PDEs}. Cambridge University Press; 2014.

\bibitem{stochastic_nematic}
Brze\'{z}niak~Z, Hausenblas~E, Razafimandimby~PA. {Some results on the
  penalised nematic liquid crystals driven by multiplicative noise: weak
  solution and maximum principle}. Stochastic partial differential equations :
  analysis and computations. 2019;\hspace{0pt}7(3):417--475.

\bibitem{stochastic_nematic_LDP}
Brze\'{z}niak~Z, Manna~U, Panda~AA. {Large Deviations for Stochastic Nematic
  Liquid Crystals Driven by Multiplicative Gaussian Noise}. Potential analysis.
  2020;\hspace{0pt}53(3):799--838.

\bibitem{wang_canevari_majumdar_2019}
Wang~Y, Canevari~G, Majumdar~A. {Order reconstruction for nematics on squares
  with isotropic inclusions: a Landau-de Gennes study}. SIAM Journal on Applied
  Mathematics. 2019;\hspace{0pt}79(4):1314--1340.

\bibitem{majumdar_2012_radialhedgehog}
Majumdar~A. {The radial-hedgehog solution in Landau–de Gennes’ theory for
  nematic liquid crystals}. European journal of applied mathematics.
  2012;\hspace{0pt}23(1):61--97.

\bibitem{majumdar-2010-article}
Majumdar~A. {Equilibrium order parameters of nematic liquid crystals in the
  Landau-de Gennes theory}. European Journal of Applied Mathematics.
  2010;\hspace{0pt}21(2):181--203.

\bibitem{zarnescu-2010-article}
Majumdar~A, Zarnescu~A. Landau-de {G}ennes theory of nematic liquid crystals:
  the {Oseen--Frank} limit and beyond. Archive for Rational Mechanics and
  Analysis. 2010;\hspace{0pt}196(1):227–--280.

\bibitem{golovaty2015}
Golovaty~D, Montero~JA, Sternberg~P. {Dimension Reduction for the Landau-de
  Gennes Model in Planar Nematic Thin Films}. Journal of Nonlinear Science.
  2015;\hspace{0pt}25(6):1431--1451.

\bibitem{mottram2014introduction}
Mottram~NJ, Newton~CJ. Introduction to $\textrm{Q}$-tensor theory.
  arXiv:14093542 [cond-matsoft]. 2014;\hspace{0pt}.

\bibitem{tsakonas_brown_davidson_mottram}
Tsakonas~C, Davidson~AJ, Brown~CV, et~al. Multistable alignment states in
  nematic liquid crystal filled wells. Applied physics letters.
  2007;\hspace{0pt}90(11):111913.

\bibitem{luo-2012}
Luo~C, Majumdar~A, Erban~R. Multistability in planar liquid crystal wells. Phys
  Rev E. 2012;\hspace{0pt}85(6):061702.

\bibitem{OR_kralj_majumdar}
Kralj~S, Majumdar~A. Order reconstruction patterns in nematic liquid crystal
  wells. Proceedings of the Royal Society A: Mathematical, Physical and
  Engineering Sciences. 2014;\hspace{0pt}470(2169):20140276.

\bibitem{RK_reference}
Butcher~JC. {The numerical analysis of ordinary differential equations :
  Runge-Kutta and general linear methods}. Chichester:Wiley; 1987.

\bibitem{hairer2012triviality}
Hairer~M, Ryser~M, Weber~H. Triviality of the 2d stochastic allen-cahn
  equation. Electronic Journal of Probability. 2012;\hspace{0pt}17(39):1--14.

\bibitem{ryser2012well}
Ryser~MD, Nigam~N, Tupper~PF. On the well-posedness of the stochastic
  allen--cahn equation in two dimensions. Journal of Computational Physics.
  2012;\hspace{0pt}231(6):2537--2550.

\bibitem{gard1988introduction}
Gard~TC. Introduction to stochastic differential equations ; 1988.

\bibitem{kloeden1992stochastic}
Kloeden~PE, Platen~E, Kloeden~PE, et~al. Stochastic differential equations.
  Springer; 1992.

\bibitem{arnold1995random}
Arnold~L, Jones~CK, Mischaikow~K, et~al. Random dynamical systems. Springer;
  1995.

\bibitem{wu2018random}
Chunrong~F, Wu~Y, Zhao~H. Anticipating random periodic solutions--ii. spdes
  with multiplicative linear noise. arXiv preprint arXiv:180300503.
  2018;\hspace{0pt}.

\bibitem{Yue_invariant_measure}
Liu~W, Mao~X, Wu~Y. {The backward Euler-Maruyama method for invariant measures
  of stochastic differential equations with super-linear coefficients}. Applied
  Numerical Mathematics. 2023;\hspace{0pt}184:137--150.

\bibitem{lewis_experiments}
Lewis~AH, Garlea~I, Alvarado~J, et~al. {Colloidal liquid crystals in
  rectangular confinement: Theory and experiment}. Soft Matter.
  2014;\hspace{0pt}10(39):7865–7873.

\bibitem{virga}
Kralj~S, Rosso~R, Virga~EG. {Finite-size effects on order reconstruction around
  nematic defects}. Phy Rev E. 2010;\hspace{0pt}81(2):021702.

\bibitem{lamy_hedgehog}
Lamy~X. {Some properties of the nematic radial hedgehog in the Landau–de
  Gennes theory}. Journal of mathematical analysis and applications.
  2013;\hspace{0pt}397(2):586--594.

\bibitem{experiments}
Sofi~JA, Dhara~S. {Stability of liquid crystal micro-droplets based optical
  microresonators}. Liquid Crystals. 2019;\hspace{0pt}46(4):629–639.

\end{thebibliography}

\end{document}